\documentclass[12pt]{article}
\usepackage{a4,graphicx}
\usepackage[small,bf]{caption}

\def\lsim{\mathrel{\rlap{\lower4pt\hbox{\hskip1pt$\sim$}}
    \raise1pt\hbox{$<$}}}                
\def\gsim{\mathrel{\rlap{\lower4pt\hbox{\hskip1pt$\sim$}}
    \raise1pt\hbox{$>$}}}                

\newcommand{\vuh}{\delta{\mu}_u} 
\newcommand{\vdh}{\delta{\mu}_d} 
 
\newcommand{\vsh}{\delta{\mu}_s} 

\newcommand{\vah}{\delta{\mu}_a}
\newcommand{\vbh}{\delta{\mu}_b}

\begin{document}

\title{
\vspace{-3.25cm}
\flushright{\small ADP-14-36/T895} \\
\vspace{-0.35cm}
{\small DESY 14-220} \\
\vspace{-0.35cm}
{\small Edinburgh 2014/20} \\
\vspace{-0.35cm}
{\small Liverpool LTH 1027} \\
\vspace{-0.35cm}
{\small April 15, 2015}  \\
\vspace{0.5cm}
\centering{\Large \bf A lattice determination of Sigma -- Lambda mixing}}

\author{\large
         R. Horsley$^a$, J. Najjar$^b$, Y. Nakamura$^c$, H. Perlt$^d$, \\
         D. Pleiter$^e$, P.~E.~L. Rakow$^f$, G. Schierholz$^g$, \\
         A. Schiller$^d$, H. St\"uben$^h$ and J.~M. Zanotti$^i$ \\[1em]
         -- QCDSF-UKQCD Collaboration -- \\[1em]
        \small $^a$ School of Physics and Astronomy,
               University of Edinburgh, \\[-0.5em]
        \small Edinburgh EH9 3FD, UK \\[0.25em]
        \small $^b$ Institut f\"ur Theoretische Physik,
               Universit\"at Regensburg, \\[-0.5em]
        \small 93040 Regensburg, Germany \\[0.25em]
        \small $^c$ RIKEN Advanced Institute for
               Computational Science, \\[-0.5em]
        \small Kobe, Hyogo 650-0047, Japan \\[0.25em]
        \small $^d$ Institut f\"ur Theoretische Physik,
               Universit\"at Leipzig, \\[-0.5em]
        \small 04109 Leipzig, Germany \\[0.25em]
         \small$^e$ J\"ulich Supercomputer Centre,
               Forschungszentrum J\"ulich, \\[-0.5em]
        \small 52425 J\"ulich, Germany \\[0.25em]
        \small $^f$ Theoretical Physics Division,
               Department of Mathematical Sciences, \\[-0.5em]
        \small University of Liverpool,
               Liverpool L69 3BX, UK \\[0.25em]
        \small $^g$ Deutsches Elektronen-Synchrotron DESY, \\[-0.5em]
        \small 22603 Hamburg, Germany \\[0.25em]
        \small $^h$ Regionales Rechenzentrum, Universit\"at Hamburg, \\[-0.5em]
        \small 20146 Hamburg, Germany \\[0.25em]
        \small $^i$ CSSM, Department of Physics,
               University of Adelaide, \\[-0.5em]
        \small Adelaide SA 5005, Australia}

\date{}

\maketitle



\begin{abstract}
Isospin breaking effects in baryon octet (and decuplet) masses
are due to a combination of up and down quark mass differences
and electromagnetic effects and lead to small mass splittings.
Between the Sigma and Lambda this mass splitting is much larger,
this being mostly due to their different wavefunctions.
However when isospin is broken, there is a mixing between between
these states. We describe the formalism necessary to determine
the QCD mixing matrix and hence find the mixing angle and mass splitting
between the Sigma and Lambda particles due to QCD effects.
\end{abstract}







\section{Introduction} 

\begin{figure}[htbp]
   \begin{center}
      \includegraphics[width=6.50cm]{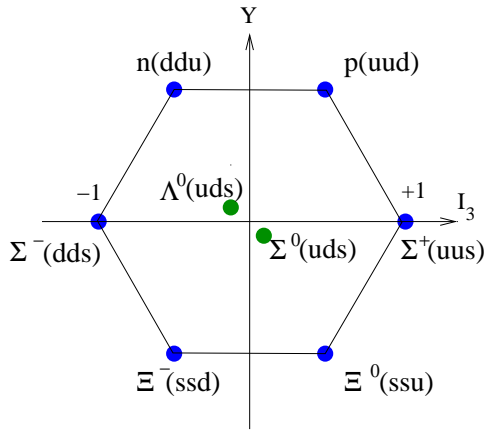}
   \end{center}
\caption{The lowest octet for the spin ${\textstyle{1\over 2}}$ baryons
         plotted in the $I_3$--$Y$ plane.}
\label{baryon_j=half_octet}
\end{figure}

Mass breaking effects in hadron octets (and decuplets)
are mainly due to a combination of quark mass differences and
electromagnetic effects, but can also sometimes have an additional
component due to mixing between the hadron states. 
In this article we consider the baryon octet
as shown in Fig.~\ref{baryon_j=half_octet} where the spin
${\textstyle{1\over 2}}$ baryons
are plotted in the $I_3$--$Y$ plane. The particles on the (outer)
ring, namely the $n(ddu)$, $p(uud)$, $\Sigma^-(dds)$, $\Sigma^+(uus)$
and $\Xi^-(ssd)$, $\Xi^0(ssu)$ all consist of combinations of $aab$ quarks
(where we use the notation of denoting a quark, $q$,
by $a$, $b$, $\ldots$ which can be the up $u$, down $d$ or strange
$s$ quark). $a$ here are the flavour doubly represented quarks,
while $b$ is the flavour singly represented quark.
For $u$ -- $d$ quark mass differences these isospin breaking effects
are small. Examples for the lowest baryon octet are the $n - p$,
$\Sigma^- - \Sigma^+$ and $\Xi^- - \Xi^0$ mass differences.
In \cite{horsley12a} we investigated the hadronic QCD
contribution to these isospin breaking splittings using lattice techniques.
In this article we extend these results to the $\Sigma^0 - \Lambda^0$ 
baryon octet masses. The method developed here for the $\Sigma^0 - \Lambda^0$
mass splitting will automatically encompass the other splittings.

The $\Sigma^0$ and $\Lambda^0$ masses%
\footnote{We use $\Sigma$ to stand for the unmixed Sigma particle
(pure isospin 1) and $\Sigma^0$ to denote the physical
Sigma particle, with mixed isospin. Similarly $\Lambda$ denotes
the pure isospin 0 state, and $\Lambda^0$ the physical Lambda
particle.}
are accurately known; from the Particle Data Group \cite{olive14a}
we have 
\begin{eqnarray}
   M_{\Sigma^0}^{\exp}  = 1.192642(24) \,\mbox{GeV}\,, \qquad  
   M_{\Lambda^0}^{\exp} = 1.115683(6) \,\mbox{GeV} \,,
\label{Sig_Lam_expt}
\end{eqnarray}
giving a mass splitting of
\begin{eqnarray}
   (M_{\Sigma^0} - M_{\Lambda^0})^{\exp} = 76.959(23) \,\mbox{MeV} \,.
\label{sig_lam_diff}
\end{eqnarray}
This is very much larger than the other mass splittings mentioned
above, which are all of the order of a few $\mbox{MeV}$. It is also
more complicated than other mass splittings as while both baryons
have the same quark content, namely $u, d, s$, most of the mass
difference is due to their different wave functions. However there will
also be additional mixing between these states. This will be apparent
when we later consider $\Sigma(ll^\prime s)$ and $\Lambda(ll^\prime s)$
where $l$ and $l^\prime$ are distinct quarks, but mass degenerate,
which already has this large mass splitting.

\begin{figure}[htbp]
   \begin{center}
      \includegraphics[width=9.00cm]
         {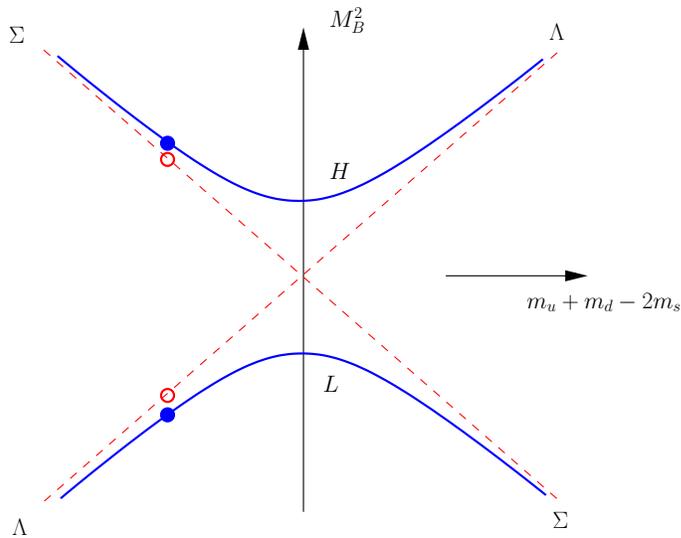}
   \end{center}
\caption{A sketch of the heavy, $H$, and light, $L$, baryon 
         $(\mbox{masses})^2$ against $m_u+m_d-2m_s$ for fixed
         $m_u - m_d$. The mass splitting between the Sigma and
         Lambda masses in the isospin limit ($m_u = m_d$) is given
         by the difference between the (red) circles; if $m_u \not= m_d$
         then there is an additional mass difference due to mixing,
         the filled (blue) circles. Further explanation of the figure
         is given in the text.} 
\label{mixing_sketch}
\end{figure}

Understanding how this mixing works will be useful for understanding
other mixing cases, such as $\eta - \eta^\prime$ or $\omega - \phi$
meson mixing, for which lattice simulations are considerably 
more difficult as there are computationally intensive disconnected
terms in the correlation function to consider,
\cite{christ10a,dudek11a,gregory11a,michael13a}. In these latter
cases a state at the centre of the octet (the pure `$\eta_8$'
octet state) mixes with a further singlet state, `$\eta_1$'.
The case here of $\Sigma^0 - \Lambda^0$ mixing is a little different
as the particles have the same quantum numbers but now lie in the
same octet (as shown in Fig.~\ref{baryon_j=half_octet}).
In Fig.~\ref{mixing_sketch}
we sketch the expected situation for the Lambda and Sigma hadrons,
plotting $M_B^2$ against $m_u + m_d - 2m_s$. The lines represent
lines of constant $m_u - m_d$, with the (red) dashed lines for
$m_u - m_d = 0$, while the (blue) lines are for $m_u - m_d \not= 0$.
The central point is the quark mass symmetric point,
when all quark masses are the same, when there is no difference
between the Lambda and Sigma masses. In the isospin limit,
when $m_u = m_d \not= m_s$ we sit at the points denoted by an
open (red) circle. The mass splitting between the Sigma and
Lambda particles is given by the vertical difference between
these points.

However if $m_u \not = m_d$ then we have mixing between
the `Lambda' and `Sigma' particles, as also depicted in the figure
by (blue) lines. The physical $\Sigma^0$ and $\Lambda^0$ masses
are now given by the (blue) filled circles. We see that there is
then an additional mass splitting.

As can also be seen from the figure, depending on the numerical
values of the quark masses, the physical $\Sigma^0$ and
$\Lambda^0$ masses can have a larger or smaller component
of the original `$\Sigma$' and `$\Lambda$' particles. 
To avoid confusion we shall call in future the lower branch
the `Light' or $L$ branch with associated mass $M_L$,
while the upper is the `Heavy' or $H$ branch with mass $M_H$.
For example in the isospin limit $m_u = m_d \equiv m_l$ we have
\begin{eqnarray}
   M_H = \left\{ \begin{array}{cc}
                    M_{\Sigma}  & m_l < m_s \\
                    M_{\Lambda} & m_l > m_s \\
                 \end{array}
         \right. \,, \qquad
   M_L = \left\{ \begin{array}{cc}
                    M_{\Lambda}  & m_l < m_s \\
                    M_{\Sigma} & m_l > m_s \\
                 \end{array}
         \right. \,.
\label{branches_mueqmd}
\end{eqnarray}
At the physical point, denoted by a $^*$, we set
\begin{eqnarray} 
   M_{\Sigma^0} = M_H^*\,, \qquad M_{\Lambda^0} = M_L^* \,.
\label{phys_Msig_Mlam}
\end{eqnarray}

In the following we denote the pure octet, i.e.\ unmixed $\Sigma$
and $\Lambda$ mass states, by the Hermitian matrix
\begin{eqnarray}
   \left( \begin{array}{cc}
             M^2_{\Sigma\Sigma}  & M^2_{\Sigma\Lambda}  \\
             M^2_{\Lambda\Sigma} & M^2_{\Lambda\Lambda} \\
          \end{array}
   \right) \,,
\label{unmixed_states}
\end{eqnarray}
while the mixed mass states will be denoted by
$M_{H}^2$, $M_{L}^2$. We determine the mixing angle,
$\theta_{\Sigma\Lambda}$, which rotates eq.~(\ref{unmixed_states})
with rotation matrix
\begin{equation}
   R = \left( \begin{array}{cc}
                 \cos\theta_{\Sigma\Lambda} 
                          & e^{i\phi_{\Sigma\Lambda}}\sin\theta_{\Sigma\Lambda}  \\
                 - e^{-i\phi_{\Sigma\Lambda}}\sin\theta_{\Sigma\Lambda} 
                          & \cos\theta_{\Sigma\Lambda} \\
              \end{array}
       \right) \,,
\label{rotation_matrix}
\end{equation}
to the diagonal form
\begin{eqnarray}
   \left( \begin{array}{cc}
             M^2_{H}  &         0    \\
                0    & M^2_{L}       \\
          \end{array}
   \right) \,,
\label{physical_states}
\end{eqnarray}
where $\phi_{\Sigma\Lambda}$ is the phase. 
Note that for the general symmetry arguments used here it does
not matter whether they are applied to the hadron mass matrix,
or some function of the mass matrix. We have chosen the quadratic
form (see section~\ref{mass_matrix_symmetries}).

Although we discuss mixing between the $\Lambda$ and
$\Sigma$ particles induced by quark mass differences,
we neglect electromagnetic effects, which will also contribute
mixing of roughly the same order of magnitude as isospin breaking 
effects. Thus we consider `pure' QCD effects only.
The method also applies to mixing of $J^P = {\textstyle{1\over 2}}^+$ 
baryons in the singly charmed sector. For $csu$ (or $csd$) baryons
the hadronic mixing will be far larger than electromagnetic effects.

Previous determinations of $\Sigma$ -- $\Lambda$ mixing include
using the quark model, e.g.\ \cite{isgur80a},
chiral perturbation theory, e.g.\ \cite{gasser82a}
and from `sum rule' methods, e.g.\ \cite{zhu98a,yagisawa01a}.

The plan of this article is as follows. In the next section,
section~\ref{SU3_flavour_expansion}, we first discuss in more detail
the calculational strategy that we employ here. In particular
as summarised in section~\ref{flavour_expan}, and discussed further
in Appendices~\ref{example_mass_matrix_symmetries} and
\ref{octet_baryon_mass_mat} we make a $SU(3)$ flavour expansion
about a point with degenerate mass $u$, $d$ and $s$ quarks.
Section~\ref{mixing_mass} then gives the $\Sigma$ -- $\Lambda$
mass mixing expansion up to NLO (i.e.\ next-to-leading order
or quadratic in the quark masses). We have actually computed
the expansion to NNLO (i.e.\ next-next-to-leading order),
but as we only use these to help to estimate systematic errors,
the complete expansions are relegated to Appendix~\ref{NNLO}.
We also show numerical simulations with two mass degenerate sea
quark masses as sufficient to determine the expansion
coefficients also for the non-degenerate quark mass case.
In section~\ref{scale_indept} we modify the expansion, to
consider ratios, rather than lattice or scale dependent quantities.
Some comments on matrix elements are given in section~\ref{matrix_els}.
Our numerical simulations are then detailed in section~\ref{num_det}
and correlation functions and determination of the expansion
coefficients are given in sections~\ref{corr_fun} and
\ref{num_coefs}, together with results for mass degenerate quarks.
Finally our results and discussion are given in section~\ref{results}.


\section{The $SU(3)$ flavour expansion}
\label{SU3_flavour_expansion}



\subsection{Mass matrix symmetries}
\label{mass_matrix_symmetries}


When all three quarks have the same mass, an $SU(3)$ transformation
$U$ on the quark fields is a symmetry of the action; it leaves 
the quark mass matrix, ${\cal M}$ unchanged. We, however,
are more interested in what happens in the case of unequal quark
masses
\begin{eqnarray}
   {\cal M} = \left( \begin{array}{ccc}
                        m_u  & 0  & 0   \\
                        0   & m_d & 0   \\
                        0   & 0   & m_s \\
                     \end{array}
              \right) \,,
\label{qm_matrix}
\end{eqnarray} 
when we make an $SU(3)$ transformation
\begin{eqnarray}
   {\cal M}^\prime = U {\cal M} U^\dagger \,.
\label{quarktrans} 
\end{eqnarray}
Although this changes the quark matrix, it does not really change 
the physical situation. The eigenvalues of ${\cal M}^\prime$ are the
same as those of ${\cal M}$, only the eigenvectors have been changed. 
Likewise, the mass spectrum of composite particles such as the 
mesons and baryons will not change, only the eigenvectors change. 

This is easiest to see if the transformation $U$ is simply 
a permutation. For example, if we interchange $m_d$ and $m_s$ 
we still get the same set of baryon masses, 
(see Fig.~\ref{baryon_j=half_octet}); all that changes is 
the names we give them. In this case, $M_n$ and $M_{\Xi^0}$ 
would be interchanged, as would $M_p$ and $M_{\Xi^+}$ and so on. 
A rotation of the quark mass matrix simply leads to a corresponding 
rotation of the baryon mass matrix, $M$, 
\begin{equation} 
   M( U {\cal M} U^\dagger ) = U M({\cal M}) U^\dagger \,.
\label{barytrans}
\end{equation} 
The $U$ matrices in eq.~(\ref{quarktrans}) belong to a $3 \times 3$ 
matrix representation of $SU(3)$, while the $U$ matrices
in eq.~(\ref{barytrans}) belong to an $8 \times 8$ representation
of the same group. 

We can see from eq.~(\ref{barytrans}) that the mass matrix 
and the $(\mbox{mass matrix})^2$ both transform in the same way
\begin{equation} 
   M^2 \to (U M U^\dagger ) (U M U^\dagger ) = U M^2 U^\dagger
\label{M2symm} 
\end{equation} 
(where, as always, $M^2$ is shorthand for $M M$). 
Therefore, as far as symmetry arguments go, it makes no 
difference whether we discuss the hadron mass matrix, or the 
mass-squared matrix. Note also that we can see from
eq.~(\ref{M2symm}) that the eigenvectors of $M$ and of $M^2$
are the same. 

We consider in future the $SU(3)$ flavour breaking
expansion of $M^2$ rather than $M$, \cite{gasser82a}.
Thus we set
\begin{eqnarray} 
   M^2 
   = \pmatrix{
       M^2_n & 0 & 0 & 0 & 0  & 0 & 0 & 0 \cr  
       0 & M^2_p  & 0 & 0 & 0  & 0 & 0 & 0 \cr  
       0 & 0 & M^2_{\Sigma^-}  & 0 & 0 & 0  & 0 & 0 \cr  
       0 & 0 & 0 & M^2_{\Sigma\Sigma} & M^2_{\Sigma\Lambda} & 0 & 0 & 0 \cr  
       0 & 0 & 0& M^2_{\Lambda\Sigma} & M^2_{\Lambda\Lambda} & 0 & 0 & 0\cr
       0 & 0 & 0 & 0 & 0 & M^2_{\Sigma^+}  & 0 & 0 \cr  
       0 & 0 & 0 & 0 & 0 & 0 & M^2_{\Xi^-}  & 0 \cr  
       0 & 0 & 0 & 0 & 0 & 0& 0  & M^2_{\Xi^0}    
            } \,.
\label{M2had_mat}
\end{eqnarray} 
The reason is that as in \cite{horsley12a} we have found again
that better numerical fits in the quark mass range considered
are obtained using the hadron mass matrix squared.

In Appendix~\ref{example_mass_matrix_symmetries} an explicit
example for the transformation $u \leftrightarrow d$ is given.


\subsection{The $\Sigma$ -- $\Lambda$ mass matrix}
\label{non-diag_mat}



\subsubsection{Derivation}


The $SU(3)$ flavour expansion classifies mass polynomials according
to the $S_3$ permutation group and the $SU(3)$ flavour group.
$S_3$ is the symmetry group of an equilateral triangle, $C_{3v}$.
This group has $3$ irreducible representations,
\cite{atkins70a}, two different singlets, $A_1$ and $A_2$ and a
doublet $E$, with elements $E^+$ and $E^-$. Some details of this group
and its representations are given in Appendix~A of \cite{bietenholz11a}.

In \cite{bietenholz11a} we classified the $10$ matrices
($N_i$, $i = 1, \ldots, 10$) which can contribute to the octet
baryon mass matrix eq.~(\ref{M2had_mat}) according to their
permutation, $S_3$ and $SU(3)$ symmetry, see Table~\ref{mat8}.
\begin{table}[htbp]
   \begin{center}
   \begin{tabular}{rrrrrrrrcc}
      $n $ & $p $ & $\Sigma^-$ & $\Sigma$ & $\Lambda$ & $\Sigma^+$
                               & $\Xi^-$ & $\Xi^0$ & $S_3$  & $SU(3)$   \\
      \hline
      1 & 1 & 1 & 1 & 1 & 1 & 1 & 1 & $A_1$ & 1                         \\
      \hline
      $-1$ & $-1$ & 0 & 0 & 0 & 0 & 1 & 1 & $E^+$ & $8_a$               \\
      $-1$ & 1 & $-2$ & 0 & 0 & 2 & $-1$ & 1 &$ E^-$ &$ 8_a$            \\
      \hline
      1 &  1 & $-2$& $-2$& 2 & $-2$& 1 & 1 & $E^+$ & $8_b$              \\
      $-1$ & 1 & 0 & \multicolumn{2}{c}{\ mix} & 0 & 1 & $-1$
                                                       & $E^-$ & $8_b$  \\
      \hline
      1 & 1 & 1 &$-3$&$-3$& 1 & 1 & 1 & $A_1$ & 27                      \\
      1 & 1 &$-2$&  3 &$-3$&$ -2$& 1 & 1 & $E^+$ & 27                   \\
      $-1$ & 1 & 0 & \multicolumn{2}{c}{\ mix} & 0 & 1 & $-1$
                                                       & $E^-$ & $27$   \\
      \hline
      1 & $-1$ & $-1$ & 0 & 0 & 1 & 1 & $-1$
        & $\phantom{^{I^X}} A_2\phantom{^{I^X}}$ & 10,${\overline{10}}$  \\
      0 & 0 & 0 & \multicolumn{2}{c}{\ mix} & 0 & 0 & 0 & $A_2$
                                                & 10,${\overline{10}}$  \\
      \hline
   \end{tabular}
   \end{center}
\caption{Mass matrix contributions for octet baryons, classified by
         permutation and $SU(3)$ symmetry. (See Table~V in
         \protect\cite{bietenholz11a}.)}
\label{mat8}
\end{table}
The compact notation of Table~\ref{mat8} gives just the diagonal elements
(the rows/columns being denoted by $n, p, \ldots$).
From Table~\ref{mat8} we see that seven of the matrices are diagonal,
they can be read off directly from the table. For example the first
row gives the $8\times 8$ matrix: $\mbox{diag}(1,1,1,1,1,1,1,1)$.
The table also contains three matrices which mix the $\Sigma$
and $\Lambda$, the fifth, eighth rows which mix at the quadratic
quark mass level and the tenth row which mixes with the cubic terms.
All the matrices are explicitly listed in
Appendix~\ref{octet_baryon_mass_mat}. Thus we write
\begin{eqnarray}
   M^2 = \sum_{i=1}^{10} \, K_i \, N_i \,,
\end{eqnarray}
where $K_i$ are some functions of the quark masses (to be determined).

We now need the three non-diagonal matrices in full. 
From Appendix~\ref{octet_baryon_mass_mat} they are $N_5$, $N_8$
and $N_{10}$. We thus have
\begin{eqnarray} 
 \matrix{ \cr \cr \cr E^- \qquad 8_b \cr \cr \cr \cr }  
  & \qquad \qquad \qquad  &  \quad
 \pmatrix{ -1 & 0 & 0 & 0 & 0 & 0 & 0 & 0 \cr
  0 & 1 & 0 & 0 & 0 & 0 & 0 & 0 \cr
  0 & 0 & 0 & 0 & 0 & 0 & 0 & 0 \cr
  0 & 0 & 0 & 0 & \frac{2}{\sqrt{3}} & 0 & 0 & 0 \cr
  0 & 0 & 0 & \frac{2}{\sqrt{3}} & 0 & 0 & 0 & 0 \cr
  0 & 0 & 0 & 0 & 0 & 0 & 0 & 0 \cr
  0 & 0 & 0 & 0 & 0 & 0 & 1 & 0 \cr
  0 & 0 & 0 & 0 & 0 & 0 & 0 & -1 } \label{mixmat8b} \\[1.2em]
 \matrix{ \cr \cr \cr  E^- \qquad 27 \cr  \cr \cr \cr }  
  &  &  
 \pmatrix{ -1 & 0 & 0 & 0 & 0 & 0 & 0 & 0 \cr
  0 & 1 & 0 & 0 & 0 & 0 & 0 & 0 \cr
  0 & 0 & 0 & 0 & 0 & 0 & 0 & 0 \cr
  0 & 0 & 0 & 0 & -\sqrt{3}  & 0 & 0 & 0 \cr
  0 & 0 & 0 & -\sqrt{3}  & 0 & 0 & 0 & 0 \cr
  0 & 0 & 0 & 0 & 0 & 0 & 0 & 0 \cr
  0 & 0 & 0 & 0 & 0 & 0 & 1 & 0 \cr
  0 & 0 & 0 & 0 & 0 & 0 & 0 & -1 } \label{mixmat27} \\[1.2em]
 \matrix{ \cr \cr \cr A_2 \ \quad 10,\overline{10} \cr  \cr \cr \cr }  
  & & \qquad 
 \pmatrix{ 0 & 0 & 0 & 0 & 0 & 0 & 0 & 0 \cr
  0 & 0 & 0 & 0 & 0 & 0 & 0 & 0 \cr
  0 & 0 & 0 & 0 & 0 & 0 & 0 & 0 \cr
  0 & 0 & 0 & 0 & -i & 0 & 0 & 0 \cr
  0 & 0 & 0 & i & 0 & 0 & 0 & 0 \cr
  0 & 0 & 0 & 0 & 0 & 0 & 0 & 0 \cr
  0 & 0 & 0 & 0 & 0 & 0 & 0 & 0 \cr
  0 & 0 & 0 & 0 & 0 & 0 & 0 & 0 } \,.
\label{mixmatA2} 
\end{eqnarray} 
We are now ready to write down the general form of the 
$\Sigma - \Lambda$ mass matrix. From Table~\ref{mat8} 
we see that the $A_1$ terms always make equal contributions
to the $\Sigma$ and $\Lambda$; and the $E^+$ terms always
make opposite contributions to the $\Sigma$ and $\Lambda$. 
From eqs.~(\ref{mixmat8b}) and (\ref{mixmat27}) we see that 
$E^-$ terms contribute a real symmetric mixing term, 
and from eq.~(\ref{mixmatA2}) that $A_2$ terms contribute 
an imaginary, antisymmetric mixing. The allowed form 
of the $\Sigma - \Lambda$ mass matrix
eq.~(\ref{unmixed_states}) is therefore
\begin{eqnarray}
   \lefteqn{\left( \begin{array}{cc}
                      M_{\Sigma \Sigma}^2 & M_{\Sigma \Lambda}^2 \\
                      M_{\Lambda\Sigma}^2 & M_{\Lambda\Lambda}^2 \\
                   \end{array} \right)}
     & &                                         \label{genform}   \\
     &=&  P_{A_1} \left( \begin{array}{cc}
                           1 & 0 \\
                           0 & 1 \\
                        \end{array} \right)
        + P_{E^+} \left( \begin{array}{cc}
                           1 & 0 \\
                           0 & -1 \\
                        \end{array} \right)
        + P_{E^-} \left( \begin{array}{cc}
                           0 & 1 \\
                           1 & 0 \\
                        \end{array} \right)
        + P_{A_2} \left( \begin{array}{cc}
                           0 & -i \\
                           i & 0 \\
                        \end{array}
                \right) \,,
                                                        \nonumber
\end{eqnarray}
where $P_G$ means a function of the quark masses with the 
symmetry $G$ under the $S_3$ permutation group. 

We can also give a permutation argument for eq.~(\ref{genform}). 
The  $\Sigma$ and $\Lambda$ form an $E$ representation of the
permutation group, with the pure $\Sigma$ even under
$u \leftrightarrow d$ and the $\Lambda$ odd. If $m_u \ne m_d$
there will be mixing between these states. Because the $\Sigma$
and $\Lambda$ have opposite behaviours under $u \leftrightarrow d$
exchange, the mass matrix for the $\Sigma - \Lambda$ system must
have the behaviour
\begin{eqnarray}
   \left( \begin{array}{cc}
             {\rm even} & {\rm odd } \\
             {\rm odd } & {\rm even} \\
          \end{array}
   \right) \,,
\label{evenodd}
\end{eqnarray}
under the operation $u \leftrightarrow d$. The possible symmetries
of the terms in the mass matrix are given by
\begin{equation}
   E \otimes E = A_1 \oplus E \oplus A_2 \,.
\end{equation}
The $A_1$ and the $E^+$ member of the $E$ doublet are even
under $u \leftrightarrow d$, so they must be responsible
for the diagonal part of the mass matrix. The mixing 
terms in the mass matrix are odd, so they must come from 
$E^-$ and $A_2$ expressions. 

From the above discussion we note that the formalism includes
the no--mixing case when $m_u = m_d$; we simply set
\begin{eqnarray}
   P_{E^-} = 0 \,, \qquad P_{A_2} = 0 \,,
\end{eqnarray}
and the upper component of eq.~(\ref{genform}), now in a diagonal form,
gives the degenerate mass of the Sigma baryons: 
$\Sigma \equiv (\Sigma^-,\Sigma^0,\Sigma^+)$ (which upon interchanging
the quarks also gives the other baryon masses on the outer ring
$N \equiv (n,p)$, $\Xi \equiv (\Xi^-,\Xi^0)$),
while the lower component gives $\Lambda$.


\subsubsection{Diagonalisation}


We now diagonalise the $2 \times 2$ $(\mbox{mass matrix})^2$ of
eq.~(\ref{genform}) giving eigenvalues
\begin{eqnarray}
   M^2_{H} &=& P_{A_1} + \sqrt{P_{E^+}^2 + P_{E^-}^2 + P_{A_2}^2}
                                                          \nonumber   \\
   M^2_{L} &=& P_{A_1} - \sqrt{P_{E^+}^2 + P_{E^-}^2 + P_{A_2}^2} \,,
\label{M2eigenval}
\end{eqnarray}
while if the eigenvectors are written as
\begin{eqnarray}
   e_{H} 
      = \left( \begin{array}{c}
                  \cos\theta_{\Sigma\Lambda}                     \\
                  e^{-i \phi_{\Sigma\Lambda}} \sin\theta_{\Sigma\Lambda} \\
               \end{array}
       \right)\,,
   \qquad 
   e_{L} 
      = \left( \begin{array}{c}
                  -e^{i \phi_{\Sigma\Lambda}} \sin\theta_{\Sigma\Lambda} \\
                   \cos\theta_{\Sigma\Lambda}                    \\
               \end{array}
        \right) \,,
\label{eigenvectors}
\end{eqnarray}
(cf eq.~(\ref{rotation_matrix})) we have 
\begin{equation}
   \tan 2\theta_{\Sigma\Lambda} = {\sqrt{P_{E^-}^2 + P_{A_2}^2} \over P_{E^+}} \,,
   \qquad 
   \tan\phi_{\Sigma\Lambda} = {P_{A_2} \over P_{E^-}} \,,
\label{mixing_angle}
\end{equation}
for the mixing angle, $\theta_{\Sigma\Lambda}$, and phase, $\phi_{\Sigma\Lambda}$. 
Note that eq.~(\ref{M2eigenval}) trivially gives the
$H$ and $L$ masses and also the mass difference
$M_{H} - M_{L}$.

Alternatively the $P_G$ coefficients have some nice links to the
$H$ and $L$ masses. $P_{A_1}$ gives the average $(\mbox{mass})^2$
\begin{equation}
   {\textstyle{1\over 2}} \left( M^2_{H} + M^2_{L} \right) 
      = P_{A_1} \,, 
\label{average_A1}
\end{equation}
while the other three coefficients contribute symmetrically to the
splitting of the two states 
\begin{equation}
   {\textstyle{1\over 2}} \left(M^2_{H} - M^2_{L} \right)
      = \sqrt{P_{E^+}^2 + P_{E^-}^2 + P_{A_2}^2} \,.
\label{diff_EpEm}
\end{equation}


\subsection{The $SU(3)$ flavour expansion}
\label{flavour_expan}


Our strategy, as discussed in detail in \cite{bietenholz11a}
is to start from a point in the quark mass plane with all three
sea quark masses equal, 
\begin{equation}
    m_u = m_d = m_s \equiv m_0 \,,
\end{equation}
and extrapolate towards the physical point,
denoted by a star, $^*$, keeping the average sea quark mass
\begin{equation}
   \bar{m} = {\textstyle{1\over 3}} (m_u + m_d + m_s)
\end{equation}
constant at the value $m_0$. As we approach the physical point,
the $u$ and $d$ quarks become lighter, but the $s$ quark becomes heavier.
Pions are decreasing in mass, but $K$ and $\eta$ increase in mass
as we approach the physical point. Keeping $\bar{m}$ constant
greatly reduces the number of mass polynomials which can occur in Taylor
expansions of physical quantities within  an $SU(3)$ multiplet.
As we are expanding about the symmetric point, it is useful to introduce
the notation
\begin{equation}
   \delta m_q \equiv m_q - \bar{m} \,,  \qquad q = u, d, s \,.
\label{delta_mq}
\end{equation}
Note that it follows from the definition that
\begin{equation}
   \delta m_u + \delta m_d + \delta m_s = 0 \,,
\label{zerosum}
\end{equation}
so we could eliminate one of the $\delta m_q$s. However we often
keep all three terms as we can then write some expressions in a more
obviously symmetrical form.

We can also generalise the $SU(3)$ flavour expansion to the case when
the mass of the valence quarks can be different to the mass of
the sea quarks, i.e.\ we leave the `unitary line'. We call this
the `partially quenched' or PQ case. To do this we introduce 
\begin{equation}
   \delta\mu_q = \mu_q - \bar{m}\,,
                 \qquad q = u, d, s \,,
\label{PQ_def}
\end{equation}
where $\mu_q$ is the valence quark mass. In distinction to the
sea quarks there is no restriction of the form eq.~(\ref{zerosum})
on the values of the valence quark masses. We give our results
in this slightly more general case and then specialise to the
unitary case $\delta\mu_q \to \delta m_q$ and then to the
physical point $\delta m_q \to \delta m_q^*$. This generalisation
will prove to be useful for the numerical determination of the
$SU(3)$ expansion coefficients.

In the following we give $SU(3)$ flavour symmetry breaking
expansions up to cubic terms in the quark's mass, i.e.\ to $O(\delta\mu_q^3)$
(in both the sea and valence quarks). We call this the `next to next to
leading order' or NNLO. However practically we shall see that the
cubic terms contribute a small amount, so we shall regard this
order as mainly being `control' on the NLO results (for which
analytic results are also given). In Table~\ref{PQpoly} we
\begin{table}[htb]
   \begin{center}
   \begin{tabular} {ccccc}
   Polynomial  & $\qquad S_3 \qquad$ & \multicolumn{3}{c}{$SU(3)$}    \\
   \hline
   $1$  & $A_1$ & $1$ &  &                              \\
   \hline \hline
   $\vuh + \vdh +\vsh$ &  $A_1$ & $1$ &  &                      \\
   \hline
   $2 \vsh - \vuh - \vdh $ &  $E^+$ &  & $8$ &                       \\
   $  \vuh - \vdh$ &  $E^-$ &  & $8$ &     \\
   \hline \hline
   $(\vuh + \vdh +\vsh)^2$ &  $A_1$ & 1 &  &                         \\
   $(\vuh + \vdh +\vsh) ( 2 \vsh - \vuh - \vdh) $ &   $E^+$ &  & $8$ &    \\
   $(\vuh + \vdh +\vsh)( \vuh - \vdh)$  & $E^-$ &  & $8$ & \\
   \hline
   $(\vsh -\vuh)^2 + (\vsh-\vdh)^2 +(\vuh-\vdh)^2 $ & $A_1$ & 1 &  &  $27$ \\
   $(\vsh -\vuh)^2 + (\vsh-\vdh)^2-2 (\vuh-\vdh)^2$ & $E^+$ & & $8$ &  $27$ \\
   $ (\vsh -\vuh)^2 -  (\vsh-\vdh)^2 $ &  $E^-$ &  & $8$ & $27$    \\
   \hline \hline
   $\delta m_u^2 + \delta m_d^2 +\delta m_s^2$ & $A_1$ & 1 &  &  $27$   \\
   \hline \hline
\end{tabular}
\end{center}
\caption{All the quark mass polynomials needed for
         partially quenched masses, classified by
         symmetry properties. The table includes entries up to
         $O(\delta\mu_q^2)$. (Table~XIV of~\cite{bietenholz11a}.)}
\label{PQpoly}
\end{table}
give the results to NLO.


\section{The $\Sigma$ -- $\Lambda$ mixing mass formula}
\label{mixing_mass}



\subsection{Expansion of the $P_G$ coefficients}


We now return to the evaluation of the $\Sigma$ -- $\Lambda$
mass matrix as discussed in section~\ref{non-diag_mat} and
demand that under all $SU(3)$ transformations
\begin{eqnarray}
   {\cal M} \to {\cal M^\prime} = U{\cal M}U^\dagger 
   \quad \Leftrightarrow \quad
   M^2({\cal M}^\prime) = UM^2({\cal M})U^\dagger \,.
\end{eqnarray} 
Physically there is no change, just a relabelling of the states.
For example $m _d \leftrightarrow m_s$ is equivalent to relabelling
$M_n \leftrightarrow M_{\Xi^0}\,, \ldots$

The most general form of the partially quenched octet baryon mass matrix,
for $1+1+1$ valence and sea quarks, up to order $\delta \mu_q^3$,
in the case where $\bar{m}$ is held constant
can now be determined. In Appendix~\ref{octet_baryon_mass_mat}
we illustrate explicitly the computation to leading order (LO) of the
$SU(3)$ flavour expansion and $\Sigma$ -- $\Lambda$ mixing.
We find that the coefficients%
\footnote{Note that $A_1$ and $A_2$ are used both for the $S_3$
representation and the expansion coefficient. Hopefully
this will cause no confusion in the following.}
in the $\Sigma$ -- $\Lambda$ mixing matrix, eq.~(\ref{genform}), are
\begin{eqnarray} 
   P_{A_1} 
      &=& M_0^2 + 3A_1 \delta\bar{\mu}
                                                          \nonumber   \\
      & & + {\textstyle{1\over 6}} B_0 
                      ( \delta m_u^2 + \delta m_d^2 + \delta m_s^2)  
          + B_1 ( \delta\mu_u^2 + \delta\mu_d^2  + \delta\mu_s^2 ) 
                                                          \nonumber   \\
      & &      + {\textstyle{1\over 4}} (B_3 + B_4)     
          \left[ (\delta\mu_s - \delta\mu_u)^2 
                + (\delta\mu_s - \delta\mu_d)^2 + (\delta\mu_u - \delta\mu_d)^2 
          \right]
                                                          \nonumber   \\
      & &      + C_0 \delta m_u \delta m_d \delta m_s 
               + 3C_1 \delta\bar{\mu}
                                (\delta m_u^2+\delta m_d^2+\delta m_s^2) 
                                                          \nonumber   \\
      & &      - 4(C_5 + C_7) \delta\mu_u \delta\mu_d \delta\mu_s 
               + {\textstyle{1\over 2}} Q_1 ( \delta\mu_s + \delta\mu_u)
                         (\delta\mu_s + \delta\mu_d)(\delta\mu_u+\delta\mu_d) 
                                                          \nonumber   \\
      & &      + {\textstyle{27\over 4}}Q_2 (\delta\mu_s - \delta\bar{\mu})
                                    (\delta\mu_u - \delta\bar{\mu})
                                    (\delta\mu_d - \delta\bar{\mu}) \,,
                                                          \nonumber   \\
   P_{E^+} 
      &=& {\textstyle{3\over 2}} A_2 ( \delta\mu_s - \delta\bar{\mu} )
                                                          \nonumber   \\
      & & + {\textstyle{1\over 2}} B_2 
                           ( 2 \delta\mu_s^2 - \delta\mu_u^2 - \delta\mu_d^2)
                                                          \nonumber   \\
      & & + {\textstyle{1\over 4}} (B_3 - B_4)  
          \left[ (\delta\mu_s - \delta\mu_u)^2 
                + (\delta\mu_s - \delta\mu_d)^2 -2 (\delta\mu_u - \delta\mu_d)^2
          \right] 
                                                          \nonumber   \\
      & & + {\textstyle{3\over 2}} C_2  (\delta\mu_s - \delta\bar{\mu})
                              (\delta m_u^2 + \delta m_d^2 + \delta m_s^2)
          + 6 (C_3 - C_4)
             ( \delta\mu_s - \delta\bar{\mu}) \delta\bar{\mu}^2 
                                                          \nonumber   \\
      & & + {\textstyle{1\over 6}} Q_3
          \left[ (\delta\mu_s - \delta\mu_u)^2 + 
                   (\delta\mu_s - \delta\mu_d)^2 
                   - 2 (\delta\mu_u - \delta\mu_d)^2 
          \right]\delta\bar{\mu}
                                                          \nonumber   \\
      & & + {\textstyle{1\over 8}} Q_4 
           (\delta\mu_s - \delta\bar{\mu})
             (\delta\mu_u^2 + \delta\mu_d^2 + \delta\mu_s^2 
              - 3\delta\bar{\mu}^2) \,,
                                                          \nonumber   \\
   P_{E^-} 
      &=& {\textstyle{\sqrt{3}\over 2}} A_2 (\delta\mu_d - \delta\mu_u) 
                                                          \nonumber   \\
      & & + {\textstyle{\sqrt{3}\over 2}} B_2 (\delta\mu_d^2 - \delta\mu_u^2)
          + {\textstyle{\sqrt{3}\over 4}} (B_3 - B_4) 
            \left[(\delta\mu_s - \delta\mu_d)^2 - (\delta\mu_s - \delta\mu_u)^2
            \right]
                                                          \nonumber   \\
      & & + {\textstyle{\sqrt{3}\over 2}} C_2 (\delta\mu_d - \delta\mu_u)
                            (\delta m_u^2 + \delta m_d^2 + \delta m_s^2) 
          + 2 \sqrt{3} (C_3 - C_4)(\delta\mu_d - \delta\mu_u)
                                      \delta\bar{\mu}^2 
                                                          \nonumber   \\
      & & + {\textstyle{1\over 8\sqrt{3}}} Q_4
             (\delta\mu_d - \delta\mu_u)
              (\delta\mu_u^2 + \delta\mu_d^2 + \delta\mu_s^2 
                - 3\delta\bar{\mu}^2 ) 
                                                          \nonumber   \\
       & & - {\textstyle{\sqrt{3}\over 2}} Q_3
                (\delta\mu_d - \delta\mu_u)
                  (\delta\mu_s - \delta\bar{\mu})\delta\bar{\mu} \,,
                                                          \nonumber   \\
    P_{A_2}
       &=& C_9 (\delta\mu_s - \delta\mu_u)(\delta\mu_s - \delta\mu_d)
                               (\delta\mu_u - \delta\mu_d)  \,,
\label{PtoO3}
\end{eqnarray} 
where
\begin{eqnarray} 
    Q_1 &\equiv & 2C_3 + C_5 + C_7
                                                          \nonumber   \\
    Q_2 &\equiv &  C_5 - C_6 + C_7 + C_8
                                                          \nonumber   \\
    Q_3 &\equiv &  4 (C_3 - C_4) + 3 (C_5 - C_7)
                                                          \nonumber   \\
    Q_4 &\equiv &  2 (C_3 - C_4) + 3 (C_5 -C_7) - 9 (C_6 + C_8) \,,
\end{eqnarray} 
and
\begin{eqnarray}
   \delta\bar{\mu} 
      \equiv {\textstyle{1\over 3}}
                       (\delta\mu_u + \delta\mu_d + \delta\mu_s) \,.
\label{mubar_def}
\end{eqnarray}
We can check that all the polynomials that occur here are 
polynomials of the appropriate symmetry from Table~\ref{PQpoly}
(i.e.\ Table~XIV of \cite{bietenholz11a}), or linear combinations
of those polynomials. For example for $E^+$ we have written
\begin{eqnarray}
   {\textstyle{1\over 2}}(2\delta\mu_s^2-\delta\mu_u^2-\delta\mu_d^2)
      &=& {\textstyle{1\over 3}}(\delta\mu_u+\delta\mu_d+\delta\mu_s)
             (2\delta\mu_s-\delta\mu_u-\delta\mu_d)
                                                                      \\
      & & + {\textstyle{1\over 6}}((\delta\mu_s-\delta\mu_u)^2
               +(\delta\mu_s-\delta\mu_d)^2 -2(\delta\mu_u-\delta\mu_d)^2) \,.
                                                           \nonumber
\end{eqnarray}
$P_{E^+}$ and $P_{E^-}$ form a doublet, i.e.\
they are related by $S_3$ symmetry, and involve the same parameters. 

Using these expansions eqs.~(\ref{M2eigenval}) and (\ref{mixing_angle})
now give the $H$ and $L$ masses, together with the mixing angle
$\theta_{\Sigma\Lambda}$ and phase $\phi_{\Sigma\Lambda}$. Note that in
the unitary limit $\delta\mu_q \to \delta m_q$ these expressions
simplify greatly
\begin{eqnarray} 
   P_{A_1} 
     &=& M^2_0 + {\textstyle{1\over 12}}(2 B_0 +12 B_1 + 9 B_3 + 9 B_4 )
                         (\delta m_u^2 + \delta m_d^2 + \delta m_s^2)  
                                                          \nonumber   \\
     & & \phantom{M^2_0} 
              + {\textstyle{1\over 4}}(4 C_0 -16 C_5 - 16 C_7 -16 Q_1 + 27 Q_2)
                          \delta m_u \delta m_d \delta m_s 
                                                          \nonumber   \\
   P_{E^+}
     &=& {\textstyle{3\over 2}} A_2 \delta m_s 
         + {\textstyle{1\over 8}}(2 B_2 +3 B_3 -3 B_4)
                   \left[ 3 \delta m_s^2 - (\delta m_u - \delta m_d)^2 \right]  
                                                          \nonumber   \\
     & &  + {\textstyle{1\over 8}}(12 C_2 + Q_4) 
                 \delta m_s ( \delta m_u^2 + \delta m_d^2 + \delta m_s^2)
                                                          \nonumber   \\
   P_{E^-}
     &=& {\textstyle{\sqrt{3}\over 2}} A_2 ( \delta m_d - \delta m_u ) 
         + {\textstyle{\sqrt{3}\over 4}} (2 B_2 +3 B_3 -3 B_4)
                 \delta m_s (\delta m_u - \delta m_d) 
                                                          \nonumber   \\
     & & + {\textstyle{1 \over 8\sqrt{3}}} ( 12 C_2 + Q_4)
                 (\delta m_u - \delta m_d)
                      ( \delta m_u^2 + \delta m_d^2 + \delta m_s^2)
                                                          \nonumber   \\
   P_{A_2} 
     &=& C_9 (\delta m_s - \delta m_u)
                    (\delta m_s - \delta m_d)(\delta m_u - \delta m_d) \,.
\label{ul_P}
\end{eqnarray}


\subsection{Mass formulae, octet hadrons, $2+1$ case}
\label{mass_2p1}


Let us now consider the equal mass valence up and down quark limit, i.e.\
\begin{eqnarray}
   \delta\mu_u = \delta\mu_d \equiv \delta\mu_l \,;
\end{eqnarray}
then $P_{E^-} = 0 = P_{A_2}$ (and $P_{A_1}$, $P_{E^+}$ simplify)
which means that $\theta_{\Sigma\Lambda} = 0$, i.e.\ there is no
$\Sigma$ -- $\Lambda$ mixing, eq.~(\ref{genform}) is already diagonal
and so
\begin{eqnarray}
   M^2_{\Sigma}  = P_{A_1} + P_{E^+}  \,, \qquad
   M^2_{\Lambda} = P_{A_1} - P_{E^+}  \,,
\label{MHL_2+1}
\end{eqnarray}
with%
\footnote{From eq.~(\ref{PtoO3}) we have
$P_{E^+} 
  = {\textstyle{3\over 2}} A_2 ( \delta\mu_s - \delta\bar{\mu} ) + \ldots
        =  A_2 ( \delta\mu_s - \delta\mu_l) + \ldots $. From
eq.~(\ref{branches_mueqmd}) (and Fig.~\ref{mixing_sketch}),
generalising to PQ quarks, eq.~(\ref{PQ_def}),
we see that with $A_2 > 0$ if $\delta\mu_l < \delta\mu_s$ then
$M_H$ describes $M_\Sigma$ while if $\delta\mu_l > \delta\mu_s$ then
$M_L$ describes $M_\Sigma$. Hence $M_\Sigma^2$ is always given by
$P_{A_1} + P_{E^+}$. Similarly $M_{\Lambda}^2$ is given by $P_{A_1} - P_{E^+}$.}
$A_2 > 0$. 

However as there is now no mixing then the mass formula must also
automatically describe the $\Sigma^+$, $\Sigma^-$ and hence all the
`outer' baryons, with flavour structure $aab$, eq.~(\ref{M2had_mat}).
Replacing $\delta\mu_l$ by $\delta\mu_a$ and  $\delta\mu_s$ by
$\delta\mu_b$ we find
\begin{eqnarray}
   M_{\Sigma}^2(aab)
      &\equiv& M_{\Sigma}^2(aa^\prime b)
                                                         \nonumber  \\
      &=& M_0^2 + A_1(2\delta\mu_a + \delta\mu_b) 
                + A_2(\delta\mu_b - \delta\mu_a)
                                                         \nonumber  \\
      & &       + {\textstyle{1\over 6}} B_0
                  (\delta m_u^2 + \delta m_d^2 + \delta m_s^2)
                                                         \nonumber  \\
      & &       + B_1(2\delta\mu_a^2+\delta\mu_b^2)
                + B_2(\delta\mu_b^2-\delta\mu_a^2) 
                + B_3(\delta\mu_b-\delta\mu_a)^2
                                                         \nonumber   \\
      & &       + C_0\delta m_u\delta m_d\delta m_s
                                                         \nonumber   \\
      & &       + [ C_1(2\delta\mu_a + \delta\mu_b) 
                    + C_2(\delta\mu_b - \delta\mu_a)
                   ](\delta m_u^2 + \delta m_d^2 + \delta m_s^2)
                                                         \nonumber   \\
      & &        + C_3(\delta\mu_a + \delta\mu_b)^3
                 + C_4(\delta\mu_a + \delta\mu_b)^2(\delta\mu_a - \delta\mu_b)
                                                         \nonumber   \\
      & &        + C_5(\delta\mu_a + \delta\mu_b)
                                     (\delta\mu_a - \delta\mu_b)^2
                 + C_6(\delta\mu_a - \delta\mu_b)^3 \,.
\label{N_2+1_pq}
\end{eqnarray}
The notation used here is meant to indicate that $a$, $a^\prime$
(and $a^{\prime\prime}$) are distinct quarks (in the baryon wave function),
but are mass degenerate, i.e.\ $\mu_a = \mu_{a^\prime}$ ($= \mu_{a^{\prime\prime}}$)%
\footnote{It should be clear from the context whether we are referring
to the Sigma particle or collectively to a particle on the outer ring
of the octet. A similar comment holds for the Lambda.}
This agrees with our previous results in \cite{horsley12a} and
\cite{bietenholz11a} (and justifies the notation for the expansion
coefficients of eq.~(\ref{PtoO3})). The valence flavour structure
of eq.~(\ref{N_2+1_pq}) then describes the broken isospin case
of $p \equiv \Sigma(uud)$, $n \equiv \Sigma(ddu)$,
$\Sigma^+ \equiv \Sigma(uus)$, $\Sigma^- \equiv \Sigma(dds)$,
and $\Xi^0 = \Sigma(ssu)$, $\Xi^- = \Sigma(ssd)$ as well
as the isospin degenerate $\Sigma^0 \equiv \Sigma(ll^\prime s)$. 
Furthermore now that we have cubic terms present, the Coleman-Glashow
mass relation \cite{coleman61a} is violated,
\begin{eqnarray}
   \lefteqn{M^2_n - M^2_p - M^2_{\Sigma^-} 
            + M^2_{\Sigma^+} + M^2_{\Xi^-} - M^2_{\Xi^0}}
     & &                                                 \nonumber   \\
     & & \hspace*{0.75in} = \,\,
         2 ( C_4 - 3 C_6) (\delta\mu_s - \delta\mu_u)(\delta\mu_s-\delta\mu_d)
                        (\delta\mu_d - \delta\mu_u)
\label{coleman_glashow}
\end{eqnarray}
(compare with eq.~(38) in \cite{bietenholz11a}). 

At the cubic level we have four new coefficients $C_3$, $C_4$, $C_5$,
$C_6$ involving the valence quarks alone, and three new coefficients
$C_0$, $C_1$, $C_2$ which involve the sea quark masses, and which drop
out for calculations on the symmetric background, $\delta m_q=0$.
Eq.~(\ref{N_2+1_pq}) assumes that $\bar{m}$,
the average sea quark mass, is held constant. A large number
of additional terms appear if that constraint is relaxed. 
If we work on single background all the sea quark terms can be absorbed
into the valence parameters; the $B_0$ and $C_0$ terms can be absorbed
into $M^2_0$; $C_1$ and $C_2$ can be absorbed into $A_1$ and $A_2$
respectively. 

Useful for numerical simulations is to take mass degenerate
$u$ and $d$ sea quarks, so we have $2+1$ flavours rather than $1+1+1$
in the generation of configurations
\begin{eqnarray}
   \delta m_u = \delta m_d \equiv \delta m_l \,,
\label{isospin_lim}
\end{eqnarray}
which (together with eq.~(\ref{zerosum})) is equivalent to the replacements
\begin{eqnarray}
   \delta m_l^2 
      \leftrightarrow
          {\textstyle{1\over 6}}(\delta m_u^2 + \delta m_d^2 + \delta m_s^2)
      \,, \qquad
   \delta m_l^3 
      \leftrightarrow
         - {\textstyle{1\over 2}}\delta m_u\delta m_d\delta m_s \,,
\label{1+1+1_sea}
\end{eqnarray}
in eq.~(\ref{PtoO3}) for $P_G$, $G = A_1$, $E^+$, $E^-$, $A_2$.

Similarly we can write down the mass of the octet Lambda baryon as
\begin{eqnarray}
   M_{\Lambda}^2(aa^\prime b)
      &=& M_0^2 + A_1(2\delta\mu_a + \delta\mu_b) 
                - A_2(\delta\mu_b - \delta\mu_a)
                                                         \nonumber  \\
      & &       + {\textstyle{1\over 6}} B_0
                  (\delta m_u^2 + \delta m_d^2 + \delta m_s^2)
                                                         \nonumber  \\
      & &       + B_1(2\delta\mu_a^2+\delta\mu_b^2)
                - B_2(\delta\mu_b^2-\delta\mu_a^2) 
                + B_4(\delta\mu_b-\delta\mu_a)^2
                                                         \nonumber   \\
      & &       + C_0\delta m_u\delta m_d\delta m_s
                                                         \nonumber   \\
      & &       + [ C_1(2\delta\mu_a + \delta\mu_b) 
                    - C_2(\delta\mu_b - \delta\mu_a)
                   ](\delta m_u^2 + \delta m_d^2 + \delta m_s^2)
                                                         \nonumber   \\
      & &        + C_3(\delta\mu_a + \delta\mu_b)^3
                 + (C_4-2C_3)(\delta\mu_a + \delta\mu_b)^2
                                        (\delta\mu_b - \delta\mu_a)
                                                         \nonumber   \\
      & &
                 + C_7(\delta\mu_a + \delta\mu_b)(\delta\mu_b - \delta\mu_a)^2
                 + C_8(\delta\mu_b - \delta\mu_a)^3 \,.
\label{Lam_2+1_pq}
\end{eqnarray}
If all three quark masses are the same then all the masses 
become degenerate,
\begin{eqnarray}
   M_{\Sigma}^2(aaa^{\prime\prime})
      \equiv M_{\Sigma}^2(aa^\prime a^{\prime\prime})
      \equiv M_{\Lambda}^2(aa^\prime a^{\prime\prime}) \,.
\label{Maaa}
\end{eqnarray}

In addition, for mass degenerate valence up and down quarks,
the expansion in eq.~(\ref{N_2+1_pq}) now incorporates
not only the mass degenerate nucleons, $p, n \equiv \Sigma(lll^{\prime\prime})$
with mass $M_N = M_\Sigma(lll^{\prime\prime})$, the Sigmas,
$\Sigma^-, \Sigma^+ \equiv \Sigma(lls)$ and
$\Sigma^0 \equiv \Sigma(ll^\prime s)$ with mass
$M_\Sigma = M_\Sigma(lls) = M_\Sigma(ll^\prime s)$
and Xis, $\Xi^-, \Xi^0 \equiv \Xi(ssl)$
with mass $M_\Xi = M_\Sigma(ssl)$, but can also be extended to incorporate
the fictitious baryon, $N_s(sss^{\prime\prime})$ with mass
$M_\Sigma(sss^{\prime\prime})$.
Furthermore the expansion in eq.~(\ref{Lam_2+1_pq}) also
incorporates not only the Lambda, $\Lambda \equiv \Lambda(ll^\prime s)$
with mass $M_\Lambda = M_\Lambda(ll^\prime s)$, but can be extended to
the fictitious baryon $\Lambda_{2sl}(ss^\prime l)$ with mass
$M_\Lambda(ss^\prime l)$. For three mass degenerate valence quarks,
we see that this mass formula reduces to the previous formula
$M_\Lambda(ll^\prime l^{\prime\prime}) = M_\Sigma(lll^{\prime\prime})$ and 
$M_\Lambda(ss^\prime s^{\prime\prime}) = M_\Sigma(sss^{\prime\prime})$.

The $\Lambda$ mass formula involves several parameters $B_4$, $C_7$ 
and $C_8$ which do not appear in the other baryon masses. We can
understand why some terms in $M_\Lambda^2$ are constrained by the other 
hadron masses, while others are independent. The partially quenched
quantities
\begin{eqnarray} 
   (X^2_N)^{PQ} 
      \equiv  {\textstyle{1\over 3}} (M^2_N + M^2_\Sigma+M^2_\Xi)
      \,, \qquad
   (X^2_\Lambda)^{PQ} 
      \equiv  {\textstyle{1\over 2}} (M^2_\Lambda + M^2_\Sigma)
\end{eqnarray} 
agree with each other exactly if $\delta\mu_s = \delta\mu_l$
(unbroken valence $SU(3)$, eq.~(\ref{Maaa})).
The quantity $(X^2_\Lambda)^{PQ} - (X^2_N)^{PQ}$ is a $27$-plet,
so it should be $O((\delta\mu_s-\delta\mu_l)^2)$ if
valence $SU(3)$ is broken. Therefore any terms in $M^2_\Lambda$ which survive,
or vanish more slowly than $(\delta\mu_s-\delta\mu_l)^2$, as
$\delta\mu_s \to \delta\mu_l$ are constrained by the other baryon masses; 
any terms which vanish like $(\delta\mu_s-\delta\mu_l)^2$ or faster
can have new independent coefficients unconnected to the other baryon masses. 
The $B_4$, $C_7$ and $C_8$ terms are the only terms in $M^2_\Lambda$
which vanish fast enough as  $\delta\mu_s \to \delta\mu_l$ to evade this
$(X^2_\Lambda)^{PQ} \to (X^2_N)^{PQ}$ constraint, as from
eqs.~(\ref{N_2+1_pq}) and (\ref{Lam_2+1_pq}) we find%
\footnote{ For completeness, we also give here the result for
$(X^2_\Lambda)^{\rm PQ} - (X^2_N)^{\rm PQ}$ in the full $1+1+1$ case,
generalising eq.~(\ref{XLam2pq-XN2pq}),
\begin{eqnarray} 
   (X^2_\Lambda)^{\rm PQ} - (X^2_N)^{\rm PQ}
      &=& {\textstyle{1\over 4}} (3 B_4 - B_3)
            (\delta\mu_u^2 + \delta\mu_d^2 + \delta\mu_s^2 
                - 3 \delta \bar \mu^2)
                                                        \nonumber \\
      & & + {\textstyle{3\over 4}} (3 C_7 -C_5 - 9 C_6 + 9 C_8 ) 
              (\delta\mu_s -\delta \bar \mu)(\delta\mu_u -\delta \bar \mu)
                                   (\delta\mu_d-\delta \bar \mu) 
                                                        \nonumber \\
      & & + {\textstyle{1\over 2}} (3 C_7 -6 C_3 - C_5) 
             (\delta\mu_u^2 + \delta\mu_d^2 + \delta\mu_s^2
                     - 3 \delta\bar{\mu}^2) \delta\bar{\mu} \,,
                                                        \nonumber
\end{eqnarray} 
where $\delta\bar{\mu}$ is defined in eq.~(\ref{mubar_def}).}
\begin{eqnarray} 
   (X^2_\Lambda)^{\rm PQ}- (X^2_N)^{\rm PQ} 
      &=& {\textstyle{1\over 6}} (3 B_4 - B_3 ) (\delta\mu_s - \delta\mu_l)^2   
                                                         \nonumber  \\
      & & + {\textstyle{1\over 18}} ( 3 C_7 -C_5 - 9 C_6 + 9 C_8)
                                      (\delta\mu_s - \delta\mu_l)^3 
                                                         \nonumber  \\
      & & + {\textstyle{1\over 9}} (3C_7 - 6C_3 -  C_5 ) 
                (\delta\mu_s + 2\delta\mu_l) (\delta\mu_s - \delta\mu_l)^2 \,.
\label{XLam2pq-XN2pq}
\end{eqnarray} 

At the $O(\delta \mu_q^2)$ level all the coefficients $M_0^2, A_i, B_i$
occur in eqs.~(\ref{N_2+1_pq}), (\ref{Lam_2+1_pq}), and so can be found
from a $2+1$ flavour partially quenched calculation. (The coefficients
are functions of $\bar{m}$, and so will not change from a $1+1+1$
simulation to a $2+1$ simulation provided that $\bar{m}$ is held
constant.) This will no longer hold at $O(\delta \mu_q^3)$, to find $P_{A_2}$
we would need to measure the other $A_2$ mass combination, which is
the Coleman--Glashow violation, eq.~(\ref{coleman_glashow}) or eq.~(38)
of \cite{bietenholz11a}, which requires $1+1+1$ valence quarks and a
more general expansion than given in eqs.~(\ref{N_2+1_pq}) and
(\ref{Lam_2+1_pq}), so we have introduced a new coefficient, $C_9$, here.

 
\section{Scale independent quantities}
\label{scale_indept}

 
\subsection{Ratios}
\label{rats}


We now restrict ourselves to giving results to NLO (which will be 
sufficient for our numerical determinations). For completeness the
full NNLO expressions are given in Appendix~\ref{NNLO}.

Numerically it is advantageous to consider scale independent quantities,
as previously discussed and used in \cite{horsley12a,bietenholz11a}.
As stated in section~\ref{flavour_expan} flavour blind (or singlet)
quantities are suitable to form both scale independent mass ratios
and to determine the scale. We denote these quantities generically
by $X_S$. One useful type can be considered as the `centre of mass'
of the multiplet. Thus for the baryon octet, one possibility is
\begin{eqnarray}
   X_N^2 &=& {\textstyle{1\over 6}}
             ( M_p^2 + M_n^2 + M_{\Sigma^+}^2 +  M_{\Sigma^-}^2
                    + M_{\Xi^0}^2 + M_{\Xi^-}^2 )
                                                        \nonumber \\
        &=& M_0^2 + ({\textstyle{1\over 6}}B_0 + B_1 + B_3)
                   (\delta m_u^2 + \delta m_d^2 + \delta m_s^2) \,.
\label{XN2_def}
\end{eqnarray}
At the physical point this has the value \cite{olive14a}
\begin{eqnarray}
   X_N^{\exp} = 1.1610\,\mbox{GeV} \,.
\label{XN_phys}
\end{eqnarray}
As discussed in \cite{bietenholz11a} flavour blind quantities,
due to the vanishing of the linear $\delta m_q$ terms,
(see eq.~(\ref{XN2_def})) remain almost constant as we approach
the physical point, so $a_N = (a_N X_N) / X_N^{\exp}$ determines
the lattice spacing $a_N(\kappa_0)$, \cite{bietenholz11a}.
(We have introduced the $N$ subscript as we are using $X_N$
to set the scale.)

We shall in future consider for the baryon octet the
dimensionless ratios
\begin{eqnarray}
   \tilde{M}^2 \equiv {M^2 \over X_N^2}
   \,, \quad \ldots \,,
\label{mass_rat}
\end{eqnarray}
and we wish to rewrite eq.~(\ref{M2eigenval}) in the form
\begin{eqnarray}
   \tilde{M}^2_{H}
      &=& \tilde{P}_{A_1} 
            + \sqrt{\tilde{P}_{E^+}^2 + \tilde{P}_{E^-}^2 + \tilde{P}_{A_2}^2}
                                                          \nonumber   \\
   \tilde{M}^2_{L} 
      &=& \tilde{P}_{A_1} 
            - \sqrt{\tilde{P}_{E^+}^2 + \tilde{P}_{E^-}^2 + \tilde{P}_{A_2}^2} \,,
\label{tildeM2eigenval}
\end{eqnarray}
and eq.~(\ref{mixing_angle}) as
\begin{equation}
   \tan 2\theta_{\Sigma\Lambda}
       = {\sqrt{\tilde{P}_{E^-}^2 + \tilde{P}_{A_2}^2} 
                       \over \tilde{P}_{E^+}} \,,
   \qquad 
   \tan\phi_{\Sigma\Lambda} = {\tilde{P}_{A_2} \over \tilde{P}_{E^-}} \,,
\label{tildemixing_angle}
\end{equation}
where
\begin{eqnarray}
   P_G \to \tilde{P}_G = { P_G \over X_N^2} \,, \quad G = A_1, E^+, E^-, A_2 \,.
\end{eqnarray}
This can be achieved by defining
\begin{equation}
   \tilde{A}_i = { A_i \over M_0^2} \,, \qquad
   \tilde{B}_i = { B_i \over M_0^2} \,,
\end{equation}
together with the replacement
\begin{eqnarray}
   B_0 \to \tilde{B}_0 &=& - 6 { B_1+B_3 \over M_0^2 } 
                          = - 6(\tilde{B}_1+\tilde{B}_3) \,.
\label{B0twi+C0twid}
\end{eqnarray}
The $\tilde{P}_G$, $G = A_1$, $E^+$, $E^-$, $A_2$ scale independent
flavour $SU(3)$ expansion coefficients are then given to NLO by
\begin{eqnarray} 
   \tilde{P}_{A_1} 
      &=& 1 + 3\tilde{A}_1 \delta\bar{\mu}
                                                          \nonumber   \\
      & &   + {\textstyle{1\over 6}}
                   \tilde{B}_0 ( \delta m_u^2 + \delta m_d^2 
                                                  + \delta m_s^2)  
               + \tilde{B}_1 ( \delta\mu_a^2 + \delta\mu_b^2  + \delta\mu_c^2 ) 
                                                          \nonumber   \\
      & &      + {\textstyle{1\over 4}} (\tilde{B}_3 + \tilde{B}_4)     
          \left[ (\delta\mu_c - \delta\mu_a)^2 
                + (\delta\mu_c - \delta\mu_b)^2 + (\delta\mu_a - \delta\mu_b)^2 
          \right] \,,
                                                          \nonumber   \\
   \tilde{P}_{E^+} 
      &=& {\textstyle{3\over 2}}
             \tilde{A}_2 ( \delta\mu_c - \delta\bar{\mu} )
                                                          \nonumber   \\
      & & + {\textstyle{1\over 2}}
               \tilde{B}_2 ( 2 \delta\mu_c^2 - \delta\mu_a^2 - \delta\mu_b^2)
                                                          \nonumber   \\
      & & + {\textstyle{1\over 4}} (\tilde{B}_3 - \tilde{B}_4)  
               \left[ (\delta\mu_c - \delta\mu_a)^2 
                + (\delta\mu_c - \delta\mu_b)^2 -2 (\delta\mu_a - \delta\mu_b)^2
               \right] \,,
                                                          \nonumber   \\
   \tilde{P}_{E^-} 
      &=& {\textstyle{\sqrt{3}\over 2}} \tilde{A}_2 (\delta\mu_b - \delta\mu_a) 
                                                          \nonumber   \\
      & & + {\textstyle{\sqrt{3}\over 2}}
                            \tilde{B}_2 (\delta\mu_b^2 - \delta\mu_a^2)
          + {\textstyle{\sqrt{3}\over 4}} (\tilde{B}_3 - \tilde{B}_4) 
            \left[(\delta\mu_c - \delta\mu_b)^2 - (\delta\mu_c - \delta\mu_a)^2
            \right] \,,
                                                          \nonumber   \\
    \tilde{P}_{A_2}
       &=& 0 \,,
\label{twidPtoO3}
\end{eqnarray} 
and
\begin{eqnarray}
   \delta\bar{\mu} \equiv {\textstyle{1\over 3}}
                       (\delta\mu_a + \delta\mu_b + \delta\mu_c)
\end{eqnarray}
(where we have written the more general $\delta\mu_a$, $\delta\mu_b$,
$\delta\mu_c$ rather than the previous $\delta\mu_u$, $\delta\mu_d$,
$\delta\mu_s$ respectively).

Similarly the changes to the baryon masses for mass degenerate
up and down quarks are relatively simple. For completeness we give
the scale independent flavour $SU(3)$ expansions
\begin{eqnarray}
   \tilde{M}_{\Sigma}^2(aab)
      &=& 1 + \tilde{A}_1(2\delta\mu_a + \delta\mu_b) 
                    + \tilde{A}_2(\delta\mu_b - \delta\mu_a)
                                             \label{MN_phys_cubic}  \\
      & &             + \tilde{B}_0\delta m_l^2
                      + \tilde{B}_1(2\delta\mu_a^2 + \delta\mu_b^2)
                      + \tilde{B}_2(\delta\mu_b^2 - \delta\mu_a^2) 
                      + \tilde{B}_3(\delta\mu_b - \delta\mu_a)^2 \,,
                                                         \nonumber 
\end{eqnarray}
and
\begin{eqnarray}
   \tilde{M}_{\Lambda}^2(aa^\prime b)
      &=& 1 + \tilde{A}_1(2\delta\mu_a + \delta\mu_b) 
            - \tilde{A}_2(\delta\mu_b - \delta\mu_a)
                                              \label{MLam_phys_cubic}  \\
      & &             + \tilde{B}_0\delta m_l^2
                      + \tilde{B}_1(2\delta\mu_a^2 + \delta\mu_b^2)
                      - \tilde{B}_2(\delta\mu_b^2 - \delta\mu_a^2) 
                      + \tilde{B}_4(\delta\mu_b - \delta\mu_a)^2 \,,
                                                         \nonumber 
\end{eqnarray}
where $\tilde{B}_0$ is given in eq.~(\ref{B0twi+C0twid}).

 
\subsection{Analytic expressions}
\label{analytic}


Finally we analytically expand out eqs.~(\ref{tildeM2eigenval}) and
(\ref{tildemixing_angle}) to NLO. On the unitary line (which is all
that we shall later need) this gives
\begin{eqnarray}
   \tan 2\theta_{\Sigma\Lambda}
     &=& {(\delta m_d - \delta m_u) \over \sqrt{3}\delta m_s} \times
                                        \label{phys_practical_angleoXN2} \\
     & & \hspace*{0.25in}
         \left[ 1 - {1 \over 3}\left( {2 \tilde{B}_2 + 3 \tilde{B}_3 
                                     - 3 \tilde{B}_4 \over \tilde{A}_2} \right)
                     { (\delta m_s - \delta m_u)
                       (\delta m_s - \delta m_d) \over \delta m_s }
        \right] \,,
                                                        \nonumber
\end{eqnarray}
and for the sum and difference, after additionally expanding further
in the masses (rather than $(\mbox{mass})^2$)
\begin{eqnarray}
   {1 \over 2} \left( \tilde{M}_{\Sigma^0} + \tilde{M}_{\Lambda^0} \right)
       &=& 1 + {1 \over 8}\left( -\tilde{B}_3 + 3\tilde{B}_4 
                                 - {\textstyle{3\over 2}} \tilde{A}_2^2 \right)
                  \left( \delta m_u^2 + \delta m_d^2 + \delta m_s^2 \right) \,,
\end{eqnarray}
and
\begin{eqnarray}
   \tilde{M}_{\Sigma^0} - \tilde{M}_{\Lambda^0}
      &=& \sqrt {3 \over 2} 
           \tilde{A}_2 \sqrt{\delta m_u^2 + \delta m_d^2 + \delta m_s^2} \times
                                              \label{MSigpmMLamXN}   \\
      & & \hspace*{0.35in}
        \left[ 1 + {3 \over 2}\left(2\tilde{B}_2+3\tilde{B}_3
                                   -3\tilde{B}_4 \over \tilde{A}_2\right)
                   {\delta m_u \delta m_d \delta m_s \over
                       \delta m_u^2 + \delta m_d^2 + \delta m_s^2 }
        \right] \,.
                                                        \nonumber
\end{eqnarray}
In the isospin limit, upon using eq.~(\ref{isospin_lim})
we again see that the mixing angle in eq.~(\ref{phys_practical_angleoXN2})
vanishes, but the $\Sigma$ -- $\Lambda$ mass difference in 
eq.~(\ref{MSigpmMLamXN}) still persists. Let us first examine
the convergence of the series. If we expand in terms of the quark
mass difference $\delta m_d - \delta m_u$ and, now generalising
eq.~(\ref{isospin_lim}) slightly, the average quark mass $\delta m_l$
where $\delta m_l$ is given by
\begin{eqnarray}
   \delta m_l = (\delta m_u + \delta m_d)/2 \,,
\end{eqnarray}
then
\begin{eqnarray}
   -{ (\delta m_s - \delta m_u)
       (\delta m_s - \delta m_d) \over 3\delta m_s }
     &=& { 3\over 2}\delta m_l + O((\delta m_d - \delta m_u)^2) \,,
                                                            \nonumber   \\
   {3\delta m_u \delta m_d \delta m_s \over
      2(\delta m_u^2 + \delta m_d^2 + \delta m_s^2) }
     &=& {1\over 2}\delta m_l + O((\delta m_d - \delta m_u)^2) \,.
\end{eqnarray}
At (or close to) the physical point $\delta m_d \approx \delta m_u$
so that in the expansion of $\tan 2\theta_{\Sigma\Lambda}$, as
compared to $\tilde{M}_{\Sigma^0} - \tilde{M}_{\Lambda^0}$ the NLO term
is a factor $\approx 3$ larger, and hence the convergence of the
$SU(3)$ symmetry flavour breaking series is expected to be worse
for the mixing angle than for the mass difference.

As an estimate of the contribution of isospin breaking to
$\Sigma$ -- $\Lambda$ mass splitting we expand in the difference
between the isospin breaking and isospin symmetric cases giving to LO
\begin{eqnarray}
   \left(\tilde{M}_{\Sigma^0} - \tilde{M}_{\Lambda^0}\right)
   - \left. \left(\tilde{M}_{\Sigma} - \tilde{M}_{\Lambda}\right) 
     \right|_{\delta m_l}
      = {1 \over 8} \, \tilde{A}_2 \,
          {(\delta m_d - \delta m_u)^2 \over |\delta m_l| } \,.
\label{SigLamIso}
\end{eqnarray}

Mass splitting formulae for the baryons on the outer ring
were given in \cite{horsley12a} (eqs.~(12) -- (15)). For example we have
\begin{eqnarray}
   \lefteqn{\tilde{M}_n - \tilde{M}_p  
              \equiv \tilde{M}_\Sigma(ddu) - \tilde{M}_\Sigma(uud)}
      & &                                           \label{Nucsplit}  \\
      &=& {\textstyle{1\over 2}} (\delta m_d - \delta m_u)
          \left[ \tilde{A}_1 -2\tilde{A}_2
                + (2\tilde{B}_1-4\tilde{B}_2 
                   - {\textstyle{3\over 2}}\tilde{A}_1^2
                   + 3\tilde{A}_1\tilde{A}_2) \delta m_l
          \right] \,.
                                                           \nonumber
\end{eqnarray}
There are several differences between the isospin splitting between
$\Sigma^0$ -- $\Lambda^0$ and $n$ -- $p$. For $\Sigma$ -- $\Lambda$
mixing from eq.~(\ref{SigLamIso}) we see that the mass splitting
starts quadratically in $(\delta m_d - \delta m_d)$ while from
eq.~(\ref{Nucsplit}) for $n$ -- $p$ the splitting is linear.
Furthermore from eq.~(\ref{MSigpmMLamXN})
we see that this difference depends principally on $\tilde{A}_2$ and
not at all on $\tilde{A}_1$. (The $\tilde{A}_1$ term has cancelled
in the unitary limit in eq.~(\ref{ul_P}).) This is completely
opposite to the mass splittings of the baryons on the `outer ring',
\cite{horsley12a} and eq.~(\ref{Nucsplit}), which depend on
$\tilde{A}_1$ as well as $\tilde{A}_2$. As $\tilde{A}_1$ is
numerically found to be much larger the result is then dominated
by this coefficient.

We use these expansions in our numerical determinations.
While for the central values of $M_{\Sigma^0}-M_{\Lambda^0}$
and $M_n - M_p\,, \ldots\,,$ it matters little whether we use
these expressions or directly use those in section~\ref{rats},
for the error (particularly of $M_n - M_p\,, \ldots\,$
but rather less so for $M_{\Sigma^0}-M_{\Lambda^0}$) the difference
depending on just one or two coefficients leads to a better
determination.


\section{Matrix elements}
\label{matrix_els}


While we are primarily interested in this article on masses,
we now make a few comments here on matrix elements.
We see from eq.~(\ref{SigLamIso}) that in masses isospin breaking
effects are second order in $\delta m_d - \delta m_u$. However,
if we look at transition amplitudes instead of masses,
the effects of the mixing angle appear at first order in
$\theta_{\Sigma\Lambda}$, i.e.\ at first order in $\delta m_d - \delta m_u$,
making an experimental determination of the mixing angle much more feasible.

It was pointed out in \cite{karl94a} that the semileptonic
decays $\Sigma^- \to \Lambda^0 e \bar{\nu}$ and
$\Sigma^+ \to \Lambda^0 e^+ \nu$ are particularly
sensitive to the $\Sigma$ -- $\Lambda$ mixing angle.
In the absence of mixing we would have 
\begin{eqnarray} 
   \Sigma^- \to \Sigma   
      && A = \sqrt{2} ( \gamma_\mu + F \gamma_\mu \gamma_5 )V_{ud}
                                                           \nonumber \\
   \Sigma^+ \to \Sigma
      && A = - \sqrt{2} ( \gamma_\mu + F \gamma_\mu \gamma_5 )V_{ud}
                                                           \nonumber \\
   \Sigma^- \to \Lambda
      && A = \sqrt{\textstyle \frac{2}{3}} 
                  D \gamma_\mu \gamma_5 V_{ud}
                                                           \nonumber \\
   \Sigma^+ \to \Lambda
      && A = \sqrt{\textstyle \frac{2}{3}} 
                  D \gamma_\mu \gamma_5 V_{ud} \,,
\end{eqnarray} 
where $A$ is the amplitude, $F$ and $D$ are the axial $SU(3)$ couplings
and $V_{ud} \sim \cos\theta_C$ the appropriate CKM matrix element.
There are two important points to  note about these amplitudes.
First, the $\Sigma^- \to \Lambda$ amplitude is equal to the
$\Sigma^+ \to \Lambda$  amplitude, while the $\Sigma^- \to \Sigma$
has the opposite sign to the $\Sigma^+ \to \Sigma$. 
Secondly, the $\Sigma \to \Lambda$ amplitudes are 
purely axial, while the $\Sigma \to \Sigma$ amplitudes 
have a large vector contribution. 

If we now introduce mixing as defined in eq.~(\ref{eigenvectors})
\begin{eqnarray}
   \Lambda^0 = -\sin \theta_{\Sigma\Lambda} \Sigma
                  + \cos \theta_{\Sigma\Lambda}  \Lambda \,,
\end{eqnarray}
we have 
\begin{eqnarray} 
   \!\!\! \Sigma^+ \to \Lambda^0   
      && A = \left\{  \sqrt{2} \gamma_\mu \sin \theta_{\Sigma\Lambda} 
             + \left( \sqrt{\textstyle \frac{2}{3}} D \cos \theta_{\Sigma\Lambda}
             + \sqrt{2} F \sin \theta_{\Sigma\Lambda} \right) 
                 \gamma_\mu \gamma_5 \right\} V_{ud}
                                                                     \\
  \!\!\! \Sigma^- \to \Lambda^0   
      && A = \left\{ - \sqrt{2} \gamma_\mu \sin \theta_{\Sigma\Lambda} 
             + \left( \sqrt{\textstyle \frac{2}{3}} D \cos \theta_{\Sigma\Lambda}
             - \sqrt{2} F \sin \theta_{\Sigma\Lambda} \right) 
                  \gamma_\mu \gamma_5 \right\} V_{ud} \,,
                                                           \nonumber 
 \end{eqnarray} 
for the transition amplitudes to the physical (mixed) $\Lambda^0$. 

We see two effects which might be experimentally measurable
at levels of order several percent. Firstly, the $\Sigma \to  
\Lambda$ amplitudes have acquired a small vector component, 
which should change the angular distributions of 
the decay products. Secondly, the interference between the 
$D$ and $F$ components of the amplitudes works in opposite
directions in the two cases. After correcting for phase
space differences, we should see that the total
$\Sigma^+ \to \Lambda^0$ decay rate 
is enhanced, while the $\Sigma^- \to \Lambda^0$
is suppressed by mixing. Both effects are first order 
in the mixing, and so first order in $m_d-m_u$, and so
they should be much more significant than the effect
of mixing on the hadron masses. 

In principle mixing effects of this sort appear in all 
decays of the $\Sigma^0$ and $\Lambda^0$, and all decays
with a $\Sigma^0$ or $\Lambda^0$ in the decay products. 
All the semileptonic decays effected by the $\Sigma$ -- $\Lambda$
mixing are also discussed in \cite{karl94a}.

 
\section{Determination of the expansion coefficients}
\label{num_det}


From eq.~(\ref{genform}) we see that we need to find the
$2\times 2$ mass matrix. We see from eqs.~(\ref{phys_practical_angleoXN2}),
(\ref{MSigpmMLamXN}) that to determine
$\theta_{\Sigma\Lambda}$, $M_{\Sigma^0}$, $M_{\Lambda^0}$ to LO we need to find
the $\tilde{A}_2$ coefficient; to NLO also $\tilde{B}_2$, $\tilde{B}_3$
and $\tilde{B}_4$. We also need, of course, $\delta m_u^*$, $\delta m_d^*$,
$\delta m_s^*$, i.e.\ a determination of the physical point. As apparent
from section~\ref{mass_2p1}, they can in principle all be determined
from $2+1$ simulations of the $\Sigma$ and $\Lambda$ masses.
However we have in addition also determined the 
off-diagonal matrix elements of the $2\times 2$ mass matrix
eq.~(\ref{genform}) for some PQ quark masses with
$\delta\mu_a \not= \delta\mu_b \not= \delta\mu_c$.

Numerical simulations have been performed using $n_f = 2+1$ $O(a)$
improved clover fermions \cite{cundy09a} at $\beta = 5.50$ and 
mainly on $32^3\times 64$ lattice sizes, \cite{bietenholz11a}.
Errors given here are statistical (using $\sim O(1500)$ configurations)
later together with an estimate of the systematic errors.

Once the $SU(3)$ flavour degenerate sea quark mass, $m_0$,
is chosen, subsequent sea quark mass points $m_l$, $m_s$ are then
arranged in the various simulations to keep $\bar{m}$
($ = m_0$) constant. This ensures that all the expansion coefficients
given previously do not change. In \cite{bietenholz11a} it was seen
that a linear fit provides a good description of the numerical data
on the unitary line over the relatively short distance from
the $SU(3)$ flavour symmetric point down to the physical pion mass.
This proved useful in helping us in choosing the initial point
on the $SU(3)$ flavour symmetric line to give a path that reaches
(or is very close to) the physical point.

The bare quark masses (both valence $\mu_q$ and unitary
$\mu_q \to m_q$) in lattice units are given by
\begin{eqnarray}
   \mu_q = {1 \over 2} 
            \left ({1\over \kappa_q} - {1\over \kappa_{0c}} \right)
            \qquad \mbox{with} \quad q = l, s, 0, a, b \,,
\label{kappa_bare}
\end{eqnarray}
and where vanishing of the quark mass along the $SU(3)$ flavour
symmetric line determines $\kappa_{0c}$. We denote the $SU(3)$ flavour
symmetric kappa value, $\kappa_0$, as being the initial point on the path
that leads to the physical point. This is given in eq.~(\ref{kappa_bare})
with $q = 0$ and replacing $\mu_0$ by $m_0$. Keeping 
$\bar{m} = \mbox{constant} = m_0$ then gives
\begin{equation}
   \delta\mu_q = {1 \over 2}
                 \left( {1 \over \kappa_q} - {1 \over \kappa_0}
                 \right) \,.
\label{delta_muq_kappa}
\end{equation}
We see that $\kappa_{0c}$ has dropped out of eq.~(\ref{delta_muq_kappa}),
so we do not need its explicit value here. While the choice of
partially quenched quark masses is not restricted, along the
unitary line the quark masses are restricted and we have
\begin{eqnarray}
   \kappa_s = { 1 \over { {3 \over \kappa_0} - {2 \over \kappa_l} } } \,.
\label{kappas_mbar_const}
\end{eqnarray}
So a given $\kappa_l$ determines $\kappa_s$ here. The $SU(3)$ flavour
symmetric $\kappa_0$ value chosen here for this action was found
to be $\kappa_0 = 0.12090$ \cite{bietenholz11a}. The constancy of
flavour singlet quantities along the unitary line to the physical
point \cite{bietenholz11a} leads directly from $X_N$ to an estimation
of the lattice spacing here of $a_N(\kappa_0 = 0.12090) \sim 0.079\,\mbox{fm}$.

 
\subsection{Correlation functions}
\label{corr_fun}


The wave functions (operators) used to determine the hadron masses
are all taken to be Jacobi smeared. For the $\Sigma(abc)$ and
$\Lambda(abc)$ we have
\begin{eqnarray}
   {\cal B}_{\Sigma(abc)\, \alpha}(x)
      &=& {1 \over \sqrt{2}} \epsilon^{abc} \left(
            b_\alpha^a(x)\left[a^b(x)^{T_D}C\gamma_5 c^c(x)\right]
            +a_\alpha^a(x)\left[b^b(x)^{T_D}C\gamma_5 c^c(x)\right]
                                           \right) \,,
                                                             \nonumber \\
   {\cal B}_{\Lambda(abc)\, \alpha}(x)
      &=& {1 \over \sqrt{6}} \epsilon^{abc} \left(
            2c_\alpha^a(x)\left[a^b(x)^{T_D}C\gamma_5 b^c(x)\right]
                                  \right.
                                                                       \\
      & & \hspace*{0.50in}\left.
            +b_\alpha^a(x)\left[a^b(x)^{T_D}C\gamma_5 c^c(x)\right]
            -a_\alpha^a(x)\left[b^b(x)^{T_D}C\gamma_5 c^c(x)\right]
                                  \right) \,.
                                                             \nonumber
\end{eqnarray}
where $C = \gamma_2\gamma_4$ and the superscript $^{T_D}$ denotes
a transpose in Dirac space%
\footnote{The colour indices are also denoted by $a$, $b$ and $c$;
hopefully this will cause no confusion.}.
The $\Sigma$ wave function is even under the interchange
$a \leftrightarrow b$, while the $\Lambda$ wave function is odd
under this interchange.

The correlation functions (on a lattice of temporal extension $T$
and spatial volume $V_s$) are given from the correlation matrix%
\footnote{$\Gamma_{\rm unpol} = {1 \over 2}(1 + \gamma_4)$.}
\begin{eqnarray}
   C_{ij}(t)
          &=& {1 \over V_s} \, \mbox{Tr}_D \Gamma_{\rm unpol} \, \left\langle
                 \sum_{\vec{y}} {\cal B}_i(\vec{y},t)
                 \sum_{\vec{x}} \bar{\cal B}_j(\vec{x},0)
              \right\rangle
                                                             \nonumber \\
          &\propto& A_iA_j e^{-M_Lt} + B_iB_j e^{-M_Ht} \,, 
                    \qquad 0 \ll t \ll T/2 \,,
\label{correlation_matrix}
\end{eqnarray}
with $i, j = \Sigma(abc), \Lambda(abc)$.
This matrix is diagonalised, yielding $M_H$ and $M_L$.

As many of our choices of PQ valence quark masses have degenerate mass
$a$ and $b$ quarks, then as discussed in section~\ref{mass_2p1},
some simplification for the $\Sigma$ wave function is possible.
In this case we note that the Grassmann contractions lead to
$C_{\Sigma(aa^\prime b)\Lambda(aa^\prime b)} = 0$ identically, so that,
as expected, the correlation matrix eq.~(\ref{correlation_matrix}) is diagonal.
Furthermore for the outer octet baryons we can use instead
the wave function
\begin{eqnarray}
   {\cal B}_{\Sigma(aab) \, \alpha}(x)
       = \epsilon^{abc} a_\alpha^a(x)
               \left[ a^b(x)^{T_D}C\gamma_5 b^c(x) \right] \,,
\end{eqnarray}
for valence quarks $a$ and $b$. The corresponding correlation
function is
\begin{eqnarray}
   C_{\Sigma(aab)\Sigma(aab)}(t)
          &=& {1 \over V_s} \, \mbox{Tr}_D \Gamma_{\rm unpol} \, \left\langle
                 \sum_{\vec{y}} {\cal B}_{\Sigma(aab)}(\vec{y},t)
                 \sum_{\vec{x}} \bar{\cal B}_{\Sigma(aab)}(\vec{x},0)
              \right\rangle
                                                             \nonumber \\
          &\propto& A e^{-M_{\Sigma}\,t}\,, \qquad 0 \ll t \ll T/2 \,.
\end{eqnarray}
This determines the $M_{\Sigma}(aab)$ masses. Considering the correlation
functions, $C_{\Sigma(aa^\prime b)\Sigma(aa^\prime b)}(t)$ and $C_{\Sigma(aab)\Sigma(aab)}(t)$
the Grassmann contractions can be shown to be equivalent so
\begin{eqnarray}
   C_{\Sigma(aab)\Sigma(aab)}(t) \propto C_{\Sigma(aa^\prime b)\Sigma(aa^\prime b)}(t) \,,
\end{eqnarray}
and so the masses are the same, $M_{\Sigma(aab)} = M_{\Sigma(aa^\prime b)}$,
as indicated in in eq.~(\ref{N_2+1_pq}).
Similarly when all the quark masses are degenerate, then
$C_{\Sigma(aaa^{\prime\prime})\Sigma(aaa^{\prime\prime})}(t) \propto 
C_{\Lambda(aa^\prime a^{\prime\prime})\Lambda(aa^\prime a^{\prime\prime})}(t)$
or $M_{\Sigma(aaa^{\prime\prime})} = M_{\Lambda(aa^\prime a^{\prime\prime})}$ as expected.

Fitting to eq.~(\ref{tildeM2eigenval}) then determines the
$\tilde{A}$, $\tilde{B}$ coefficients.
Together with a knowledge of the physical (and unitary) quark masses,
$\delta m_u^*$, $\delta m_d^*$, $\delta m_s^*$, this leads to
an evaluation of the physical $\Sigma^0$ and $\Lambda^0$ masses,
see eq.~(\ref{phys_Msig_Mlam}).


\subsection{Numerical results for the expansion coefficients}
\label{num_coefs}


Although simulations between the $SU(3)$ flavour symmetric point and
the physical point are in principle enough to determine the expansion
coefficients, in practice it is advantageous to increase
the range to try to determine the NLO terms more reliably
(i.e.\ with reduced error bars). However we also hope that the
$SU(3)$ flavour breaking expansion developed here remains valid.
As we see later in this section, for the $\Sigma$ -- $\Lambda$
splitting there is a further constraint. This leads to a choice of
valence quark masses in the range
$|\delta\mu_a|+|\delta\mu_b|+|\delta\mu_c| \lsim 0.2$.
This translates to nucleon masses of $\lsim 2\, \mbox{GeV}$, so
roughly the physical baryon masses lie in the middle
of our fit range.
The corresponding pion mass range is from about
$800$ down to $200\,\mbox{MeV}$, the $SU(3)$ flavour
symmetric pion lying at about $420\,\mbox{MeV}$.

In order to determine these $\tilde{A}$ and $\tilde{B}$ coefficients,
additional PQ masses have been determined on the set of gauge configurations
that have all three sea quark masses equal, i.e.\ at the $SU(3)$
flavour symmetric point $\kappa_0 = 0.12090$. For these particular
masses $\delta m_l = 0 = \delta m_s$ automatically. These masses
are a mixture of masses with three distinctive valence quark masses
(so we have mixing and $H$ and $L$ masses), together with two
mass degenerate quark data, when there is no mixing. 
Thus we now make a simultaneous fit to eq.~(\ref{tildeM2eigenval})
using the available data: the unitary data from
\cite{bietenholz11a} (the $32^3\times 64$ lattice data for $M_N$, 
$M_\Lambda$, $M_\Sigma$, $M_\Xi$ in Table~XXII) together with some
lighter quark mass data on a $48^3\times 96$ lattice.
Specifically we have used $23$ valence quark masses
on the $32^3\times 64$ lattice with
$(\kappa_l, \kappa_s) = (0.12090, 0.12090)$, four valence
quark masses on each of the $32^3\times 64$ lattice ensembles 
$(0.12104, 0.12062)$, $(0.121095, 0.120512)$ and $(0.121145, 0.120413)$
and a further four on the $48^3\times 96$ lattice ensemble with
$(0.121166, 0.120371)$. All the fit data used are given in
Appendix~\ref{pq_mass_tables}.

There are two LO ($\tilde{A}$) and four NLO ($\tilde{B}$)
coefficients to determine. Thus we have a six parameter
fit for the fit functions in eq.~(\ref{tildeM2eigenval}). It
was found advantageous to preserve the identity of the $\Sigma$
and $\Lambda$ particles whenever possible, so for the mass
degenerate PQ results, eqs.~(\ref{MN_phys_cubic}) and
(\ref{MLam_phys_cubic}) were used.  In Table~\ref{all_parms}
\begin{table}[!htb]
   \begin{center}
      \begin{tabular}{rr}
         $\tilde{A}_1$ & 10.17(12)    \\
         $\tilde{A}_2$ & 1.849(124)   \\
         \hline
         $\tilde{B}_1$ & 13.71(4.19)   \\
         $\tilde{B}_2$ & -20.02(4.70)  \\
         $\tilde{B}_3$ & -4.125(5.742)\\
         $\tilde{B}_4$ & -30.63(5.97) \\
      \end{tabular}
   \end{center}
\caption{Fit results for LO and NLO expansion coefficients.}
\label{all_parms}
\end{table}
we give the results of this fit with bootstrap errors.
With our normalisation for the expansion coefficients
all the numbers are $\sim O(10)$, except $\tilde{A}_2$ which is 
rather smaller. The (MINUIT) fit used gave $\chi^2/\mbox{dof} \sim 38/60
\sim 0.6$ per degree of freedom.

Two simple plots which illustrate the fit results are first
the completely mass degenerate case (when as discussed previously in
section~\ref{corr_fun} all outer baryon, $\Sigma$ and $\Lambda$ 
masses are the same), which may be illustrated by defining
\begin{eqnarray}
   S_{\Sigma\Lambda}
     \equiv \tilde{M}_{\Sigma}^2(aaa^{\prime\prime})
     = 1 + 3\tilde{A}_1\delta\mu_a + 3\tilde{B}_1\delta\mu_a^2 \,.
\label{S}
\end{eqnarray}
Secondly we can consider the `symmetric' difference case
(between $\Sigma$ and $\Lambda$) by setting
\begin{eqnarray}
   D^{\rm sym}_{\Sigma\Lambda}
      &\equiv& { \tilde{M}_{\Sigma}^2(aab) - \tilde{M}_{\Lambda}^2(aa^\prime b) 
          - \tilde{M}_{\Sigma}^2(bba) + \tilde{M}_{\Lambda}^2(bb^\prime a)
          \over
          4(\delta\mu_b - \delta\mu_a) }
                                                             \nonumber \\
      &=& \tilde{A}_2 + \tilde{B}_2(\delta\mu_a+\delta\mu_b) \,.
\label{Dsym}
\end{eqnarray}
(Again in these expressions and elsewhere $a^\prime$, $a^{\prime\prime},\, \ldots$
are mass degenerate but distinct quarks.) At this order
$D^{\rm sym}_{\Sigma\Lambda}$ is just a function of $\delta\mu_a+\delta\mu_b$;
at higher orders (see Appendix~\ref{NNLO}, eq.~(\ref{Dsym_NNLO})) there
are terms $\propto \delta\mu_a-\delta\mu_b$. Note that the choice for
$D^{\rm sym}_{\Sigma\Lambda}$ tends to suppress them (and indeed eliminates
them at NLO); this was the reason for the choice of this
`symmetric derivative'.

\begin{figure}[!htb]
   \begin{center}
      \includegraphics[width=9.00cm]
         {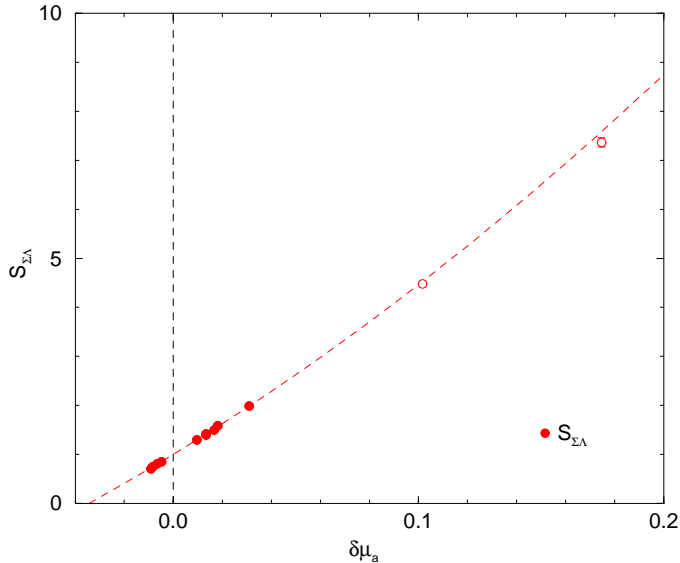}
   \end{center}
\caption{$S_{\Sigma\Lambda}$ versus $\delta\mu_a$
         ($S_{\Sigma\Lambda}$ is defined in eq.~(\protect\ref{S})),
         together with a fit also given in eq.~(\protect\ref{S}).
         Points used in the fit are denoted by filled circles
         (those outside the fit range are given by open circles).}
\label{S_plot}
\end{figure}

For the $S_{\Sigma\Lambda}$ we have the results shown in Fig.~\ref{S_plot}.
For $S_{\Sigma\Lambda}$, the fit is very good and as indicated this
could be easily extended to larger quark masses. As mentioned before
$\tilde{A}_1$ is the relevant coefficient for mass splittings on
the outer baryon ring.

\begin{figure}[!htb]
   \begin{center}
      \includegraphics[width=9.00cm]
         {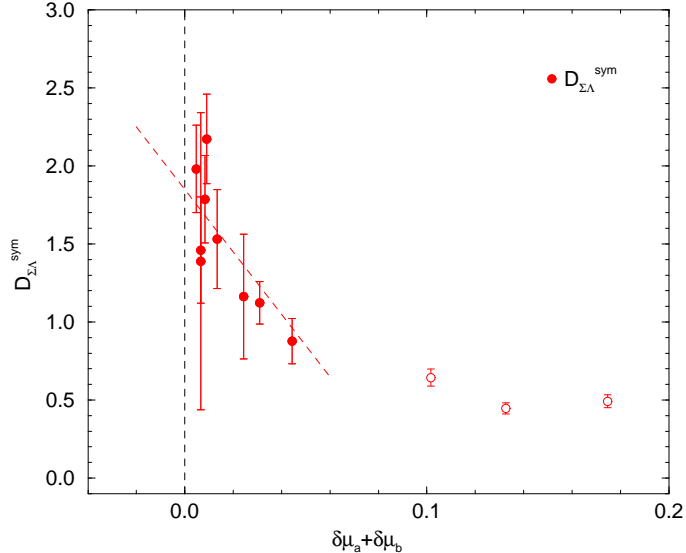}
   \end{center}
\caption{$D^{\rm sym}_{\Sigma\Lambda}$ versus $\delta\mu_a+\delta\mu_b$,
         ($D^{\rm sym}_{\Sigma\Lambda}$ is defined in
         eq.~(\protect\ref{Dsym})), together with the fit
         also given in eq.~(\protect\ref{Dsym}).
         The same notation as in Fig.~\protect\ref{S_plot}.}
\label{D_plot}
\end{figure}

In Fig.~\ref{D_plot} we plot $D^{\rm sym}_{\Sigma\Lambda}$
against $\delta\mu_a+\delta\mu_b$. We see that the data is
not linear in $\delta\mu_a+\delta\mu_b$. (As explained
before we would not expect the data in this plot to lie
on a unique curve due to the possible presence in the fit
of terms proportional to $\delta\mu_a-\delta\mu_b$.
However due to the choice of $D^{\rm sym}_{\Sigma\Lambda}$
deviations should be small.) However despite this the plot has a
sharp increase as the quark mass is reduced, indicating a
possible non-polynomial behaviour there. As this is related
to the $\Sigma$--$\Lambda$ mass splitting, this necessitates
the restricted fit region, as compared to Fig.~\ref{S_plot}.
(It should however also be noted that the unitary quark masses
have $|\delta m_a| \lsim 0.01$.)

The reason for this behaviour is due to spin--spin interaction
between the quarks. It is known (e.g.\ \cite{griffiths})
that in quark models the mass splittings are partially due
to the QCD  spin-spin interaction between the quarks.
From the Dirac equation we know that the magnetic moment of a
fermion $\propto 1/m_a$, this holds in QCD too,
for the chromomagnetic moment, which might suggest a spin--spin
interaction of the form $\propto 1/(m_am_b)$. This has
also recently been proposed in \cite{yang14a}.

 
\subsection{The physical point}
\label{physical_pt}


By considering the equivalent pseudoscalar $SU(3)$ flavour breaking
mass expansion as for the baryon octet and matching to the 
pseudoscalar meson masses gives $\delta m_u^*$, $\delta m_d^*$,
$\delta m_s^*$. Again note that by considering the outer ring of the
pseudoscalar octet, provided that the average quark mass
$\bar{m}$ is held constant, the expansion coefficients
can be determined from partially quenched $2+1$ flavour simulations
rather than $1+1+1$ flavour expansions. This was discussed in
\cite{horsley12a} (and in particular the subtraction of QED
effects) and we just quote the result of the analysis here,
as given in Table~\ref{qm_physical}. 
\begin{table}[!htb]
   \begin{center}
      \begin{tabular}{ccc}
         $\delta m_u^*$ & $\delta m_d^*$ & $\delta m_s^*$    \\
         \hline
          -0.01140(3) & -0.01067(3) & 0.02207(4)  \\
      \end{tabular}
   \end{center}
\caption{Results for the bare quark mass in lattice units at the
         physical point, slightly updated from \protect\cite{horsley12a}.}
\label{qm_physical}
\end{table}
To cover uncertainties in electromagnetic effects arising from
violations of Dashen's theorem, we assign a relative error
$\sim 15\%$ to the splitting $\delta m_d^* - \delta m_u^*$, \cite{horsley12a}.


\subsection{Comparison with `fan' plots}
\label{comparison_fan}

\begin{figure}[!htb]
   \begin{center}
      \includegraphics[width=12.00cm]
 {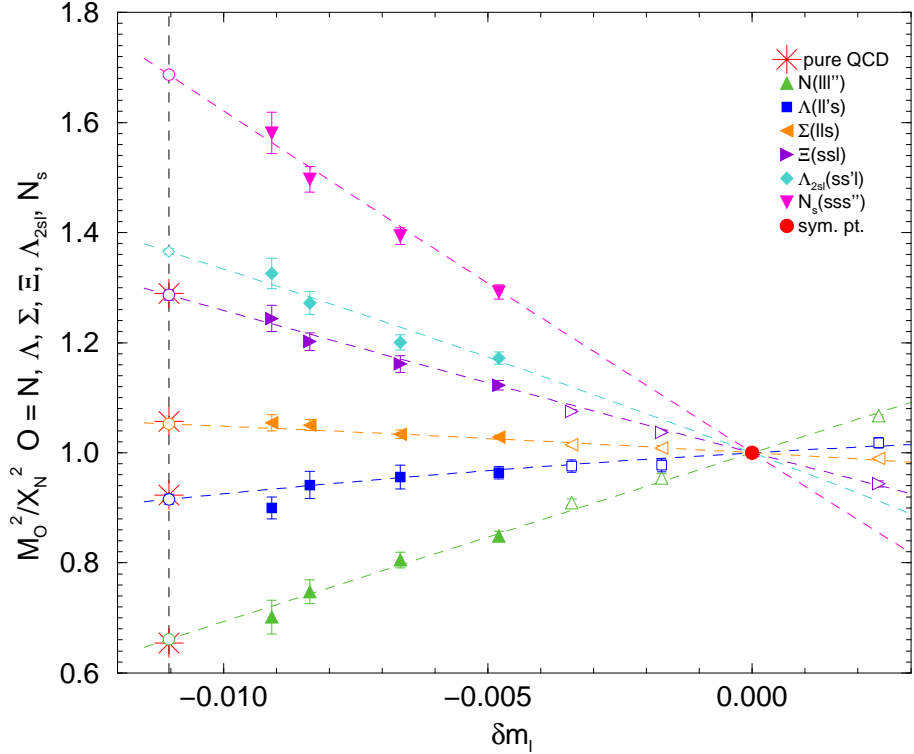}
   \end{center}
\caption{The baryon `fan' plot for the `$\Sigma$' and `$\Lambda$' type
         particles $\tilde{M}_O^2$ ($O = N$, $\Lambda$, $\Sigma$, $\Xi$,
         $\Lambda_{2sl}$, $N_s$) versus $\delta m_l$.
         Filled up triangles, squares, left triangles, right triangles,
         diamonds and down triangles are the $N(lll^{\prime\prime})$,
         $\Lambda(ll^{\prime}s)$, $\Sigma(lls)$, $\Xi(ssl)$,
         $\Lambda_{l2s}(ss^{\prime}l)$ and $N_s(sss^{\prime\prime})$ results
         respectively using $32^3\times 64$ sized lattices.
         The common symmetric point is the filled circle.
         The open up triangles, left triangles,
         right triangles, down-triangles are from comparison
         $24^3\times 48$ sized lattices (and not used in the fits here).
         The vertical dashed line from eq.~(\protect\ref{delta_ml_star})
         is the $n_f = 2+1$ pure QCD physical point, with the open
         circles being the numerically determined pure QCD hadron mass ratios
         for $2+1$ quark flavours. For comparison, the stars represent
         the average of the $(\mbox{mass})^2$ of
         $M_N^{*\,2}(lll^{\prime\prime}) = (M_n^{\exp\,2}(ddu) + M_p^{\exp\,2}(uud))/2$,
         $M_\Lambda^{*\,2}(lls) = M_{\Lambda^0}^{\exp\,2}(uds)$,
         $M_\Sigma^{*\,2}(lls)
                 = (M_{\Sigma^-}^{\exp\,2}(dds) + M_{\Sigma^+}^{\exp\,2}(uus))/2$
         and
         $M_\Xi^{*\,2}(ssl) 
                 = (M_{\Xi^-}^{\exp\,2}(ssd) + M_{\Xi^0}^{\exp\,2}(ssu))/2$.}
\label{b5p50_dml_mNmLamaaboXN2-jnt_N+L}
\end{figure}

We now compare the fit results with the mass values along the
unitary line, i.e.\ which describe the evolution of the 
baryon masses along a path from the $SU(3)$ symmetric point down
to the physical point in the isospin degenerate limit,
i.e.\ $m_u = m_d \equiv m_l$. For this comparison we take the
physical quark mass, in lattice units, from Table~\ref{qm_physical} as
\begin{eqnarray}
   \delta m_l^* \equiv (\delta m_u^* + \delta m_d^*)/2 = -0.01103(2) \,.
\label{delta_ml_star}
\end{eqnarray}
In Fig.~\ref{b5p50_dml_mNmLamaaboXN2-jnt_N+L}
we show the `fan' plot for all the `$\Sigma$' and `$\Lambda$' type particles.
We have $N(lll^{\prime\prime}) [= \Lambda_{3l}(ll^\prime l^{\prime\prime})]$,
$\Lambda(ll^\prime s)$, $\Sigma(lls)$, $\Xi(ssl)$, $\Lambda_{2sl}(ss^\prime l)$
and $N_s(sss^{\prime\prime}) [= \Lambda_{3s}(ss^\prime s^{\prime\prime}]$.
($N_s(sss^{\prime\prime})$ and $\Lambda_{2sl}(ss^\prime l)$ are fictitious
baryons, but provide additional useful data for the fits.)
As this is the diagonal case there is no mixing and from
eq.~(\ref{MHL_2+1}) the fit is given by $\tilde{M}_N^2 = P_{A_1} + P_{E^+}$,
$\tilde{M}_\Lambda^2 = P_{A_1} - P_{E^+}$. We find good agreement
with the expected results.

It can easily be seen (`ruler test') that the fits are dominated by
the LO in the $SU(3)$ flavour symmetry breaking expansion.
Given the fit results, this is not so surprising, as for the
unitary results we have a maximum quark mass given by
$|\delta m_l| \sim 0.01$, which is rather small (certainly in 
comparison with many of the PQ masses used) and indicates that
at least in the region we are interested in the low order
$SU(3)$ flavour breaking expansion describes the data well.

For completeness we give here the values at the $2+1$ QCD physical
point (open circles in Fig.~\ref{b5p50_dml_mNmLamaaboXN2-jnt_N+L})
of
$\tilde{M}^{*\,2}_N = 0.6612(58)$,
$\tilde{M}_{\Lambda}^{*\,2} = 0.9155(89)$,
$\tilde{M}^{*\,2}_\Sigma = 1.052(4)$,
$\tilde{M}^{*\,2}_\Xi = 1.286(9)$,
$\tilde{M}^{*\,2}_{\Lambda_{2sl}} = 1.365(5)$ and
$\tilde{M}^{*\,2}_{N_s} = 1.687(6)$.
For a comparison to these values, the stars in
Fig.~\ref{b5p50_dml_mNmLamaaboXN2-jnt_N+L}
represent the average of the squared experimental masses of the
appropriate particles, as defined in the figure caption.

 
\section{Results and Conclusions}
\label{results}


We now give results for the QCD contribution to the baryon masses and
their splittings.


\subsection{Outer ring of the baryon octet}
\label{outer_results}


We first discuss the masses on the outer ring of the baryon octet
using the physical quark masses given in Table~\ref{qm_physical}
and the expansion coefficients as given in Table~\ref{all_parms}.

Possible sources of systematic errors are discussed in Appendix~A
of \cite{horsley12a} as coming from the following: finite lattice volume,
convergence of the $SU(3)$ flavour symmetry breaking expansion,
the path to the physical point and finite lattice spacing
(to which we refer the reader). As the data set used has not 
changed greatly, the systematic errors are little effected,
so we use the same methods giving similar results as determined there.
For the mass ratios $\tilde{M}$ we find estimates of systematic
errors of $\sim 1\%$ for finite volume, $\sim 1\%$ for the
flavour symmetry expansion (it is also apparent from
Fig.~\ref{b5p50_dml_mNmLamaaboXN2-jnt_N+L} that in the region 
we are interested in curvature effects are very small),
$\sim 4\%$ as the chosen $\kappa_0$ and hence the trajectory in
the $m_s$ -- $m_l$ plane does not quite go through the
physical point, while the systematic errors arising from a
finite lattice spacing are small. 

We find the results for the masses and splittings of
Table~\ref{hadron_masses_av}.
\begin{table}[h]
\begin{center}

\begin{minipage}{0.45\textwidth}

   \begin{center}
      \begin{tabular}{llll}
         particle      &        & exp\,[GeV] & result\,[GeV] \\
         \hline
         $M_p$         & $uud$  & 0.9383 &  0.9427(41)(40)   \\
         $M_n$         & $ddu$  & 0.9396 &  0.9454(40)(40)   \\
         $M_{\Sigma^+}$  & $uus$  & 1.1894 &  1.1874(23)(50)   \\
         $M_{\Sigma^-}$  & $dds$  & 1.1974 &  1.1947(22)(51)   \\

         $M_{\Xi^0}$    & $ssu$  & 1.3149 &  1.3145(49)(56)   \\
         $M_{\Xi^-}$    & $ssd$  & 1.3217 &  1.3191(48)(56)   \\
      \end{tabular}
   \end{center}

\end{minipage}\hspace*{0.10\textwidth}
\begin{minipage}{0.45\textwidth}

   \begin{center}
      \begin{tabular}{ll}
         splitting      & result\,[MeV]      \\
         \hline
         $M_n - M_p$    & 2.70(15)(11)(40)    \\
         $M_{\Sigma^-} - M_{\Sigma^+}$  
                        & 7.27(22)(31)(109)   \\
         $M_{\Xi^-} - M_{\Xi^0}$
                        & 4.57(19)(19)(68)    \\
      \end{tabular}
   \end{center}

\end{minipage}
\caption{Left panel: Baryon masses on the outer ring of the octet.
         The second column gives the quark content,
         while the third column, `exp',
         gives the experimental masses from \protect\cite{olive14a}.
         The last column, `result', gives the result from
         this work. The first error is the statistical error,
         while the second is the total systematic error (in quadrature).
         $X_N^{\exp}$ from eq.~(\protect\ref{XN_phys})
         has been used to convert to $\mbox{GeV}$.
         Right panel: Baryon mass splittings on the outer ring.
         The third error is due to possible violations in Dashen's theorem,
         section~\ref{physical_pt}.}
\label{hadron_masses_av}
\end{center}
\end{table}
For the splittings, rather than using eq.~(\ref{tildeM2eigenval})
directly (i.e.\ the results of the left panel of
Table~\ref{hadron_masses_av}) we use the expressions in section~\ref{analytic}.
As discussed there, for the central values it makes little difference,
however the error is now better determined.
For the baryons on the outer ring of the octet the central
values (both for masses and mass splittings) are in
agreement with previous results, \cite{horsley12a}.
Note that we are not trying to compare the mass splittings with
the experimental values, due to electromagnetic effects
(not considered here).


\subsection{$\Sigma$ -- $\Lambda$ mixing}
\label{sig_lam_mix}


We now turn to the result for $\Sigma$ -- $\Lambda$ mixing.
In Table~\ref{sig_lam_masses_av} we give the $\Sigma^0$ and $\Lambda^0$
\begin{table}[h]
\begin{center}

   \begin{center}
      \begin{tabular}{llll}
         particle      &        & exp\,[GeV] & result\,[GeV] \\
         \hline
         $M_{\Sigma^0}$  & $uds$  & 1.1926 &  1.1910(23)(51)  \\
         $M_{\Lambda^0}$ & $uds$  & 1.1157 &  1.1109(54)(47)  \\
      \end{tabular}
   \end{center}
\caption{$\Sigma^0$ and $\Lambda^0$ masses.
         The same notation as for the left panel of 
         Table~\ref{hadron_masses_av}.}
\label{sig_lam_masses_av}
\end{center}
\end{table}

\noindent
masses. The $\Sigma^0$ -- $\Lambda^0$ mass difference is
\begin{eqnarray}
   M_{\Sigma^0} - M_{\Lambda^0} = 79.44(7.37)(3.37)\,\mbox{MeV} \,.
\end{eqnarray}
(The same discussion for the determination of the errors
as for the previous results, section~\ref{outer_results}, also
holds here.) This is to be compared with the experimental result,
eq.~(\ref{sig_lam_diff}) of $76.96(2) \,\mbox{MeV}$.
As both particles have the same quark content
(and are uncharged) we do not expect much electromagnetic
contribution. Between the LO and NLO result there is only a few percent 
difference. Furthermore taking the difference
between the $M_{\Sigma^0} - M_{\Lambda^0}$ mass splitting in
Table~\ref{sig_lam_masses_av} and $M_\Sigma^*(lls)-M_\Lambda^*(ll^\prime s)$
(i.e.\ the isospin limit) gives a tiny contribution
due to isospin breaking, consistent with zero and which
our present results are not precise enough to reliably estimate.

For the mixing angle we find
\begin{eqnarray}
   \tan 2\theta_{\Sigma\Lambda} = 0.0123(45)(25) \,,
\end{eqnarray}
which, as anticipated, gives a very small angle,
$\theta_{\Sigma\Lambda} \sim 0.006(3)\,\mbox{rads} \lsim 1^o$. 
Comparing with e.g.\ a quark model result \cite{isgur80a}
gives $\theta_{\Sigma\Lambda} \sim 0.01\,\mbox{rads}$ which is compatible
with our result.

We note that the LO value of $\tan 2\theta_{\Sigma\Lambda}$
from eq.~(\ref{phys_practical_angleoXN2}) is $\sim 0.0191$ so
in this case with our determined $\tilde{A}$ and $\tilde{B}$
values for the $SU(3)$ flavour breaking expansion, there is 
some reduction in the value of the angle when going to NLO.
However in distinction to the $\Sigma^0$ -- $\Lambda^0$ mass
difference the non-leading term is now much larger.
This is because numerically
$(\delta m_s - \delta m_u)(\delta m_s - \delta m_d) / 3\delta m_s |^* \sim
0.0166$ to be compared with
$3\delta m_u \delta m_d \delta m_s /
2(\delta m_u^2 + \delta m_d^2 + \delta m_s^2)|^* \sim 0.0056$, 
which as expected from the discussion in section~\ref{analytic}
is a factor $3$ smaller. Thus the $SU(3)$ symmetry flavour breaking 
expansion for the mixing angle in eq.~(\ref{phys_practical_angleoXN2})
appears less convergent than for the  $\Sigma^0$ -- $\Lambda^0$ mass
difference, eq.~(\ref{MSigpmMLamXN}). In order to account for this,
we have increased the relative systematic error associated with
the flavour symmetry expansion to $\sim 15\%$.


\subsection{Conclusions}


In this article we have extended our earlier work describing
the QCD contribution to isospin breaking effects in baryon masses
\cite{horsley12a} to now also include states with the same
quantum numbers, in this case the $\Sigma^0$ and $\Lambda^0$,
and their isospin mixing. This gives a complete description of
the $SU(3)$ flavour symmetry expansion of the (baryon) octet.
As an example we have numerically investigated  $\Sigma^0$ -- $\Lambda^0$
mixing. While the mass splitting is reasonably well determined,
to determine reliably the mixing angle will require a better determination
of the $SU(3)$ symmetry flavour breaking expansion. An accurate
determination of the mixing angle may be useful in baryonic semileptonic
decays, see section~\ref{matrix_els}. Further work in these directions
is in progress.


\section*{Acknowledgements}


The numerical configuration generation (using the BQCD lattice
QCD program \cite{nakamura10a}) and data analysis 
(using the Chroma software library \cite{edwards04a}) was carried out
on the IBM BlueGene/Q using DIRAC 2 resources (EPCC, Edinburgh, UK),
the BlueGene/P and Q at NIC (J\"ulich, Germany), the
SGI ICE 8200 and Cray XC30 at HLRN (The North-German Supercomputer
Alliance) and on the NCI National Facility in Canberra, Australia
(supported by the Australian Commonwealth Government).
This investigation has been supported partly 
by the EU Grants No. 227431 (Hadron Physics2) and No. 283826
(Hadron Physics3). JN was partially supported by EU grant 228398
(HPC-EUROPA2). HP is supported by DFG Grant No. SCHI 422/9-1.
JMZ is supported by the Australian Research Council Grant
No. FT100100005 and DP140103067. We thank all funding agencies.


\clearpage

\appendix

\section*{Appendix}


\section{Mass matrix symmetries -- an example}
\label{example_mass_matrix_symmetries}


To illustrate the transformations of the hadron mass
matrices with an explicit example, let 
us write out in full the symmetry matrices for the 
transformation $u \leftrightarrow d$. A $3 \times 3$ 
$SU(3)$ matrix which exchanges the $u$ and $d$ quarks 
in the quark mass matrix eq.~(\ref{qm_matrix})
is (see \cite{bietenholz11a}, eq.~(128))
\begin{equation} 
   U = \exp\left\{ i \frac{\pi}{2}( \lambda_1 + \sqrt{3} \lambda_8) \right\}
     = \pmatrix{ 0 & -1 & 0 \cr -1 & 0 & 0 \cr 0 & 0 & -1 } \,.
\label{u33} 
\end{equation} 
(The minus signs ensure that $|U| =1$, as required for an $SU(3)$
matrix). If we act with this $U$ on the quark mass matrix
it simply swaps the $u$ and $d$ quark masses. 
\begin{equation} 
   U \pmatrix{ m_u & 0 & 0 \cr 0 & m_d & 0 \cr 0 & 0 & m_s } U^\dagger
    = \pmatrix{ m_d & 0 & 0 \cr 0 & m_u & 0 \cr 0 & 0 & m_s } \,.
\end{equation} 
To transform the baryon mass matrix we need an $8 \times 8$ 
matrix corresponding to eq.~(\ref{u33}). 
\begin{equation} 
   U = \exp\left\{ i \frac{\pi}{2}( \lambda_1 + \sqrt{3} \lambda_8) \right\}
     = \pmatrix{ 0 & 1 & 0 & 0 & 0 & 0 & 0 & 0 \cr 
                 1 & 0 & 0 & 0 & 0 & 0 & 0 & 0 \cr
                 0 & 0 & 0 & 0 & 0 & 1 & 0 & 0 \cr
                 0 & 0 & 0 & -1& 0 & 0 & 0 & 0 \cr
                 0 & 0 & 0 & 0 & 1 & 0 & 0 & 0 \cr
                 0 & 0 & 1 & 0 & 0 & 0 & 0 & 0 \cr
                 0 & 0 & 0 & 0 & 0 & 0 & 0 & 1 \cr
                 0 & 0 & 0 & 0 & 0 & 0 & 1 & 0  }  \,.
\label{u88} 
\end{equation} 
found by using an $8 \times 8$ set of $\lambda$ matrices
(defined in \cite{bietenholz11a}, eq.~(144)).
 
What happens to the baryon mass matrix when we rotate it with this $U$?
\begin{eqnarray} 
   \lefteqn{ U \pmatrix{
       M^2_n & 0 & 0 & 0 & 0  & 0 & 0 & 0 \cr  
       0 & M^2_p  & 0 & 0 & 0  & 0 & 0 & 0 \cr  
       0 & 0 & M^2_{\Sigma^-}  & 0 & 0 & 0  & 0 & 0 \cr  
       0 & 0 & 0 & M^2_{\Sigma\Sigma} & M^2_{\Sigma\Lambda} & 0 & 0 & 0 \cr  
       0 & 0 & 0& M^2_{\Lambda\Sigma} & M^2_{\Lambda\Lambda} & 0 & 0 & 0\cr
       0 & 0 & 0 & 0 & 0 & M^2_{\Sigma^+}  & 0 & 0 \cr  
       0 & 0 & 0 & 0 & 0 & 0 & M^2_{\Xi^-}  & 0 \cr  
       0 & 0 & 0 & 0 & 0 & 0& 0  & M^2_{\Xi^0}    
                       } U^\dagger  } && \\[1.2em] 
    &=& 
    \pmatrix{ 
       M^2_p & 0 & 0 & 0 & 0  & 0 & 0 & 0 \cr  
       0 & M^2_n  & 0 & 0 & 0  & 0 & 0 & 0 \cr  
       0 & 0 & M^2_{\Sigma^+}  & 0 & 0 & 0  & 0 & 0 \cr  
       0 & 0 & 0 & M^2_{\Sigma\Sigma} & -M^2_{\Sigma\Lambda} & 0 & 0 & 0 \cr  
       0 & 0 & 0& -M^2_{\Lambda\Sigma} & M^2_{\Lambda\Lambda} & 0 & 0 & 0\cr
       0 & 0 & 0 & 0 & 0 & M^2_{\Sigma^-}  & 0 & 0 \cr  
       0 & 0 & 0 & 0 & 0 & 0 & M^2_{\Xi^0}  & 0 \cr  
       0 & 0 & 0 & 0 & 0 & 0& 0  & M^2_{\Xi^-}    
            } \,. \nonumber
 \end{eqnarray} 
The $n$ and $p$ switch masses, as do the $\Sigma^-$ and $\Sigma^+$ 
and the $\Xi^0$ and $\Xi^-$, all as expected when $u \leftrightarrow d$. 
In the central block, which tells us about the $\Sigma^0 \Lambda^0$ 
sector, we see that the diagonal entries are unchanged; the 
off-diagonal entries have their sign flipped. This is just what 
should happen under  $u \leftrightarrow d$; the eigenvalues
(masses of the two states) will be the same, but the mixing angle 
will be reversed, $\theta_{\Sigma\Lambda} \to -\theta_{\Sigma\Lambda}$.


\section{The octet baryon mass matrix}
\label{octet_baryon_mass_mat}



\subsection{The outer octet baryon masses}


Here we discuss the mass matrix for partially quenched 
octet baryons in more detail than we could in the body of the paper. 
The arguments given here are similar to those given in 
section B.4 of \cite{bietenholz11a} for the meson mass matrix,
and in section 4.1 for the partially quenched decuplet mass formula. 

If we have a diagonal quark mass matrix, strangeness, `upness'
and `downness' are all conserved quantum numbers. There are therefore 
only ten non-zero entries in the $8 \times 8$ octet mass matrix, 
namely the eight diagonal entries, and the two entries corresponding 
to $\Sigma$ -- $\Lambda$ mixing. $\Sigma$ --  $\Lambda$ mixing
is permitted because both baryons have the same flavour content
($uds$); any other mixing would violate flavour conservation. 

Since there are ten non-zero entries, we can express the mass
matrix in terms of a basis of ten $8 \times 8$ matrices. 
In \cite{bietenholz11a} we classified these ten matrices 
according to their symmetries; see Table~\ref{mat8}. 
Seven of the matrices are diagonal, they can be read off directly from
the table. The table also contains three matrices which mix the 
$\Sigma$ and $\Lambda$. 

In \cite{bietenholz11a} we did not specify the mixing, as we were
concentrating on the case of unbroken isospin symmetry, where there
is no mixing. We now list the basis matrices $N_i$ in full, including the
three non-diagonal matrices. 

\begin{footnotesize}
\begin{eqnarray} 
   N_1 = \pmatrix{  1 & 0 & 0 & 0 & 0 & 0 & 0 & 0 \cr
  0 & 1 & 0 & 0 & 0 & 0 & 0 & 0 \cr
  0 & 0 & 1 & 0 & 0 & 0 & 0 & 0 \cr
  0 & 0 & 0 & 1 & 0 & 0 & 0 & 0 \cr
  0 & 0 & 0 & 0 & 1 & 0 & 0 & 0 \cr
  0 & 0 & 0 & 0 & 0 & 1 & 0 & 0 \cr
  0 & 0 & 0 & 0 & 0 & 0 & 1 & 0 \cr
  0 & 0 & 0 & 0 & 0 & 0 & 0 & 1 } 
 && 
 N_2 = \pmatrix{ -1 & 0 & 0 & 0 & 0 & 0 & 0 & 0 \cr
  0 &-1 & 0 & 0 & 0 & 0 & 0 & 0 \cr
  0 & 0 & 0 & 0 & 0 & 0 & 0 & 0 \cr
  0 & 0 & 0 & 0 & 0 & 0 & 0 & 0 \cr
  0 & 0 & 0 & 0 & 0 & 0 & 0 & 0 \cr
  0 & 0 & 0 & 0 & 0 & 0 & 0 & 0 \cr
  0 & 0 & 0 & 0 & 0 & 0 & 1 & 0 \cr
  0 & 0 & 0 & 0 & 0 & 0 & 0 & 1 }  \nonumber \\[1.2em]
 N_3 = \pmatrix{  -1 & 0 & 0 & 0 & 0 & 0 & 0 & 0 \cr
  0 & 1 & 0 & 0 & 0 & 0 & 0 & 0 \cr
  0 & 0 & -2& 0 & 0 & 0 & 0 & 0 \cr
  0 & 0 & 0 & 0 & 0 & 0 & 0 & 0 \cr
  0 & 0 & 0 & 0 & 0 & 0 & 0 & 0 \cr
  0 & 0 & 0 & 0 & 0 & 2 & 0 & 0 \cr
  0 & 0 & 0 & 0 & 0 & 0 &-1 & 0 \cr
  0 & 0 & 0 & 0 & 0 & 0 & 0 & 1 } 
 && 
 N_4 = \pmatrix{  1 & 0 & 0 & 0 & 0 & 0 & 0 & 0 \cr
  0 & 1 & 0 & 0 & 0 & 0 & 0 & 0 \cr
  0 & 0 &-2 & 0 & 0 & 0 & 0 & 0 \cr
  0 & 0 & 0 &-2 & 0 & 0 & 0 & 0 \cr
  0 & 0 & 0 & 0 & 2 & 0 & 0 & 0 \cr
  0 & 0 & 0 & 0 & 0 &-2 & 0 & 0 \cr
  0 & 0 & 0 & 0 & 0 & 0 & 1 & 0 \cr
  0 & 0 & 0 & 0 & 0 & 0 & 0 & 1 }  \nonumber \\[1.2em]
 N_5 =  \pmatrix{ -1 & 0 & 0 & 0 & 0 & 0 & 0 & 0 \cr
  0 & 1 & 0 & 0 & 0 & 0 & 0 & 0 \cr
  0 & 0 & 0 & 0 & 0 & 0 & 0 & 0 \cr
  0 & 0 & 0 & 0 & \frac{2}{\sqrt{3}} & 0 & 0 & 0 \cr
  0 & 0 & 0 & \frac{2}{\sqrt{3}} & 0 & 0 & 0 & 0 \cr
  0 & 0 & 0 & 0 & 0 & 0 & 0 & 0 \cr
  0 & 0 & 0 & 0 & 0 & 0 & 1 & 0 \cr
  0 & 0 & 0 & 0 & 0 & 0 & 0 & -1 }  && 
 N_6 = \pmatrix{  1 & 0 & 0 & 0 & 0 & 0 & 0 & 0 \cr
  0 & 1 & 0 & 0 & 0 & 0 & 0 & 0 \cr
  0 & 0 & 1 & 0 & 0 & 0 & 0 & 0 \cr
  0 & 0 & 0 & -3& 0 & 0 & 0 & 0 \cr
  0 & 0 & 0 & 0 & -3& 0 & 0 & 0 \cr
  0 & 0 & 0 & 0 & 0 & 1 & 0 & 0 \cr
  0 & 0 & 0 & 0 & 0 & 0 & 1 & 0 \cr
  0 & 0 & 0 & 0 & 0 & 0 & 0 & 1 } 
 \label{matlist} \\[1.2em] 
 N_7 = \pmatrix{  1 & 0 & 0 & 0 & 0 & 0 & 0 & 0 \cr
  0 & 1 & 0 & 0 & 0 & 0 & 0 & 0 \cr
  0 & 0 &-2 & 0 & 0 & 0 & 0 & 0 \cr
  0 & 0 & 0 &  3& 0 & 0 & 0 & 0 \cr
  0 & 0 & 0 & 0 & -3& 0 & 0 & 0 \cr
  0 & 0 & 0 & 0 & 0 &-2 & 0 & 0 \cr
  0 & 0 & 0 & 0 & 0 & 0 & 1 & 0 \cr
  0 & 0 & 0 & 0 & 0 & 0 & 0 & 1 } &&
 N_8 = \pmatrix{ -1 & 0 & 0 & 0 & 0 & 0 & 0 & 0 \cr
  0 & 1 & 0 & 0 & 0 & 0 & 0 & 0 \cr
  0 & 0 & 0 & 0 & 0 & 0 & 0 & 0 \cr
  0 & 0 & 0 & 0 &\!\!\!\!\! -\sqrt{3}  & 0 & 0 & 0 \cr
  0 & 0 & 0 &\!\! -\sqrt{3}  & 0 & 0 & 0 & 0 \cr
  0 & 0 & 0 & 0 & 0 & 0 & 0 & 0 \cr
  0 & 0 & 0 & 0 & 0 & 0 & 1 & 0 \cr
  0 & 0 & 0 & 0 & 0 & 0 & 0 & -1 } \nonumber  \\[1.2em]
 N_9 = \pmatrix{  1 & 0 & 0 & 0 & 0 & 0 & 0 & 0 \cr
  0 &-1 & 0 & 0 & 0 & 0 & 0 & 0 \cr
  0 & 0 &-1 & 0 & 0 & 0 & 0 & 0 \cr
  0 & 0 & 0 &  0& 0 & 0 & 0 & 0 \cr
  0 & 0 & 0 & 0 & 0 & 0 & 0 & 0 \cr
  0 & 0 & 0 & 0 & 0 & 1 & 0 & 0 \cr
  0 & 0 & 0 & 0 & 0 & 0 & 1 & 0 \cr
  0 & 0 & 0 & 0 & 0 & 0 & 0 &-1 } &&
 N_{10} = 
 \pmatrix{ 0 & 0 & 0 & 0 & 0 & 0 & 0 & 0 \cr
  0 & 0 & 0 & 0 & 0 & 0 & 0 & 0 \cr
  0 & 0 & 0 & 0 & 0 & 0 & 0 & 0 \cr
  0 & 0 & 0 & 0 & -i & 0 & 0 & 0 \cr
  0 & 0 & 0 & i & 0 & 0 & 0 & 0 \cr
  0 & 0 & 0 & 0 & 0 & 0 & 0 & 0 \cr
  0 & 0 & 0 & 0 & 0 & 0 & 0 & 0 \cr
  0 & 0 & 0 & 0 & 0 & 0 & 0 & 0 }\nonumber  
\end{eqnarray}
\end{footnotesize}
These matrices are orthogonal, in the sense 
${\rm Tr}[ N_i N_j ] = 0$ if $i \ne j$. 
 
We can write the (mass matrix)$^2$ in terms of the basis matrices
\begin{equation} 
   M^2 = \sum_i K_i N_i \,.
\end{equation} 
This expansion is completely general. 
The coefficients $K_i$ could be functions of the pseudoscalar 
meson masses if we are doing chiral perturbation theory, 
but in our case we will use polynomials of the bare quark masses.
The symmetries of the coefficients must match the symmetries of the 
$N_i$ matrices, for example if the matrix has symmetry $A_1$ or 
$E^+$ its coefficient must be even under $m_u \leftrightarrow m_d$, 
if it has symmetry $A_2$ or $E^-$ it must be odd under this interchange. 

We find the coefficients $K_i$ by making 
all possible $SU(3)$ rotations on the quark mass matrix, 
and asking {\it Mathematica} to find the most general 
coefficients $K_i$ which lead to a $(\mbox{mass matrix})^2$
which transforms like eq.~(\ref{M2symm}). Once we know the $K_i$ 
we can then read off the individual baryon masses. 

At first order in $\delta \mu_q$ we are only allowed singlet
and octet matrices, so our $M^2$ matrix has to have the form 
\begin{equation} 
   M^2 = \sum_{i=1}^5 K_i N_i \,,
\label{K5} 
\end{equation}  
with only five terms. 

When we put this in the computer, we find that at first order,
partially quenched, the most general form of the $K_i$ 
consistent with eq.~(\ref{M2symm}) is
\begin{eqnarray} 
   K_1 &=& M^2_0 + a_1 ( \vuh  + \vdh  + \vsh  ) \nonumber \\
   K_2 &=& a_{8_a} ( 2 \vsh - \vuh - \vdh ) \nonumber \\
   K_3 &=& a_{8_a} ( \vuh - \vdh ) \label{K5init} \\
   K_4 &=&  a_{8_b} ( 2 \vsh - \vuh - \vdh ) \nonumber \\
   K_5 &=& 3 a_{8_b} ( \vuh - \vdh )\,. \nonumber  
\end{eqnarray} 
Much of this could be anticipated on general grounds. 
The form of the polynomials can be read off from 
Table~\ref{PQpoly}. Since $N_2$ and $N_3$ are part
of the same representation, we know that $K_2$ 
and $K_3$  are not independent; they must both be 
proportional to the same coefficient. Likewise, 
$K_4$ and $K_5$ must share a coefficient. 
The only slightly non-trivial features in eq.~(\ref{K5init}) are
the proportionality factors relating $K_3$ and $K_5$ 
to $K_2$ and $K_4$, (factors of $1$ and $3$). 
These have to be found by considering a symmetry operation
that mixes $N_2$ with $N_3$, and $N_4$ with $N_5$. Examples 
of such operations are the interchanges $\vdh \leftrightarrow \vsh$  
or $\vuh \leftrightarrow \vsh$, or the cyclic operation
$\vuh \to \vdh \to \vsh \to \vuh$.  

We are not quite finished; there is one extra constraint 
coming from partial quenching. If we calculate the neutron 
mass from eq.~(\ref{K5init}) we have 
\begin{eqnarray}
   \lefteqn{ 
    M_n^2 = (M^2)_{11} = K_1 - K_2 - K_3 + K_4 - K_5 } \\
        &=& M_0^2 + \vuh (a_1 - 4 a_{8_b})
            + \vdh (a_1 + 2 a_{8_a} + 2 a_{8_b})
            + \vsh (a_1 - 2 a_{8_a} + 2 a_{8_b}) \,. \nonumber 
 \end{eqnarray} 
However, we know that although the neutron can depend in a symmetric 
manner on all the sea quark masses, there is no way it can have 
any information about the mass of the partially quenched 
valence $s$ quark, so the final term should not occur. 
We remove this unwanted term by imposing the constraint 
\begin{equation} 
   a_1 - 2 a_{8_a} + 2 a_{8_b} = 0 \,,
\label{absent} 
\end{equation} 
leaving 
\begin{equation} 
   M_n^2 = M_0^2 + a_1 ( 2 \vdh + \vuh) - 4 a_{8_b} ( \vuh - \vdh) \,.
\end{equation} 
Finally, we define new parameters
\begin{equation} 
   A_1 \equiv a_1 , \qquad A_2 \equiv -4 a_{8_b} \,,
\label{newparam} 
\end{equation} 
simply to tidy up the result, 
\begin{equation} 
   M_n^2 = M_0^2 + A_1 ( 2 \vdh + \vuh) +A_2( \vuh - \vdh) \,.
\end{equation} 

The constraint eq.~(\ref{absent}) does not only remove the unphysical
term from the neutron mass formula, it automatically
does the same for all the outer baryons, giving them all
a mass formula independent of the absent valence quark mass: 
\begin{equation} 
   M^2(aab) = M^2_0+ A_1 ( 2 \vah + \vbh) + A_2 (\vbh - \vah) \,.
\label{first}
\end{equation} 

In terms of the new parameters eq.~(\ref{newparam}) the final 
expressions for the $K_i$ are 
\begin{eqnarray} 
    K_1 &=& M^2_0 + A_1 \, ( \vuh  + \vdh  + \vsh  )
                                                       \nonumber  \\
    K_2 &=& {\textstyle{1\over 4}}(2 A_1 -A_2)\, (2 \vsh -\vuh -\vdh)
                                                       \nonumber  \\
    K_3 &=& {\textstyle{1 \over 4}}(2 A_1 - A_2) \, (\vuh - \vdh) 
 \label{K5final} \\
    K_4 &=& - {\textstyle{1 \over 4}} A_2 \, (2 \vsh - \vuh - \vdh)
                                                       \nonumber  \\
    K_5 &=& - {\textstyle{3 \over 4}} A_2 \, ( \vuh - \vdh ) \,.
                                                       \nonumber 
\end{eqnarray} 

At higher order we proceed in the same way, finding the analogue
of eq.~(\ref{K5init}) by considering all possible rotations of the quark
matrix, and then the analogue of eq.~(\ref{K5final}) by imposing 
the partially quenched constraint that the absent valence quark 
can not appear in the mass formula for $M^2(aab)$. Of course
at higher order more of the $K_i$ appear; at quadratic order 
the $27$-plet enters, and we need $K_1$ to $K_8$; at cubic or higher
order, all ten $K_i$ appear. Also, the expressions for each 
$K_i$ coefficient become longer, as can be seen 
from Table~\ref{PQpoly}. It would be difficult to 
carry out the calculation by hand, but with the help of a computer we can 
find all the $K_i$, and thus the complete $M^2$ matrix. 


\subsection{The $\Sigma$ -- $\Lambda$ mass matrix}


In this paper we are primarily interested in 
the $\Sigma$ -- $\Lambda$ sector. Let us concentrate 
on the $2 \times 2$ block of $M^2$ responsible for 
these two `central' baryons. From eq.~(\ref{matlist})
we read off 
\begin{eqnarray}
   \pmatrix{ M_{\Sigma \Sigma}^2 & M_{\Sigma \Lambda}^2 \cr
    M_{\Lambda\Sigma}^2 & M_{ \Lambda\Lambda }^2 }
       = \pmatrix{ K_1 -2 K_4 -3 K_6 + 3 K_7
                 & {\textstyle{2 \over \sqrt{3}}} K_5 
                    - {\textstyle{\sqrt{3}} K_8 - i K_{10}} \cr 
                   {\textstyle{2 \over \sqrt{3}}} K_5 
                    - {\textstyle{\sqrt{3}}} K_8 + i K_{10}
                 &  K_1 +2 K_4 -3 K_6 - 3 K_7 }  \,.
\end{eqnarray}
A tidier way to write this is to split the matrix up 
according to the behaviour of the various terms under
the permutation group, eq.~(\ref{genform})
\begin{equation} 
   \pmatrix{ M^2_{\Sigma \Sigma} & M^2_{\Sigma \Lambda} \cr
    M^2_{\Lambda\Sigma} & M^2_{ \Lambda\Lambda } }
       = P_{A_1} \pmatrix{ 1 & 0 \cr 0 & 1 } 
         + P_{E^+} \pmatrix{ 1 & 0 \cr 0 & -1 } 
         + P_{E^-} \pmatrix{ 0 & 1 \cr 1 & 0 } 
         + P_{A_2} \pmatrix{ 0 & -i \cr i & 0 }  \,,
\end{equation} 
where $P_G$ means a function of the quark masses with the 
symmetry $G$ under the $S_3$ permutation group. 
The individual terms in this expansion are given by 
\begin{eqnarray} 
   P_{A_1} &=& K_1 - 3 K_6                              \nonumber  \\
   P_{E^+} &=& -2 K_4 + 3 K_7                                      \\
   P_{E^-} &=& {\textstyle{2 \over \sqrt{3}}} K_5 - {\textstyle{\sqrt{3}}} K_8
                                                      \nonumber  \\ 
   P_{A_2} &=& K_{10} \,.
                                                      \nonumber 
\end{eqnarray} 
In the main part of the paper we give the full cubic 
expression for the $P_G$, derived from the full results for the $K_i$.


 
\section{Scale independent quantities to NNLO}
\label{NNLO}


For completeness we list here the results of section~\ref{scale_indept}
to NNLO.
\begin{eqnarray}
   X_N^2 &=& M_0^2 + {\textstyle{1\over 6}}(B_0 + B_1 + B_3)
                   (\delta m_u^2 + \delta m_d^2 + \delta m_s^2)
                                                        \nonumber \\
        & & \phantom{M_0^2} + (C_0-C_3+3C_5)\delta m_u\delta m_d\delta m_s \,.
\label{XN2_def_NNLO}
\end{eqnarray}
\begin{eqnarray}
   C_0 \to \tilde{C}_0 = { C_3-3C_5 \over M_0^2 } 
                           = \tilde{C}_3-3\tilde{C}_5 \,.
\label{B0twi+C0twid_NNLO}
\end{eqnarray}
\begin{eqnarray} 
   \tilde{P}_{A_1} 
      &=& 1 + 3\tilde{A}_1 \delta\bar{\mu}
                                                          \nonumber   \\
      & &   + {\textstyle{1\over 6}}\tilde{B}_0 ( \delta m_u^2 + \delta m_d^2 
                                                  + \delta m_s^2)  
               + \tilde{B}_1 ( \delta\mu_a^2 + \delta\mu_b^2  + \delta\mu_c^2 ) 
                                                          \nonumber   \\
      & &      + {\textstyle{1\over 4}} (\tilde{B}_3 + \tilde{B}_4)     
          \left[ (\delta\mu_c - \delta\mu_a)^2 
                + (\delta\mu_c - \delta\mu_b)^2 + (\delta\mu_a - \delta\mu_b)^2 
          \right]
                                                          \nonumber   \\
      & &      + \tilde{C}_0 \delta m_u \delta m_d \delta m_s
               + 3\tilde{C}_1 \delta\bar{\mu}
                                (\delta m_u^2+\delta m_d^2+\delta m_s^2) 
                                                          \nonumber   \\
      & &      - 4(\tilde{C}_5 + \tilde{C}_7) 
                           \delta\mu_a \delta\mu_b \delta\mu_c
               + {\textstyle{1\over 2}} \tilde{Q}_1 ( \delta\mu_c + \delta\mu_a)
                         (\delta\mu_c + \delta\mu_b)(\delta\mu_a+\delta\mu_b) 
                                                          \nonumber   \\
      & &      + {\textstyle{27\over 4}}
                         \tilde{Q}_2 (\delta\mu_c - \delta\bar{\mu})
                                    (\delta\mu_a - \delta\bar{\mu})
                                    (\delta\mu_b - \delta\bar{\mu}) \,,
                                                          \nonumber   \\
   \tilde{P}_{E^+} 
      &=& {\textstyle{3\over 2}} 
                         \tilde{A}_2 ( \delta\mu_c - \delta\bar{\mu} )
                                                          \nonumber   \\
      & & + {\textstyle{1\over 2}}
               \tilde{B}_2 ( 2 \delta\mu_c^2 - \delta\mu_a^2 - \delta\mu_b^2)
                                                          \nonumber   \\
      & & + {\textstyle{1\over 4}} (\tilde{B}_3 - \tilde{B}_4)  
               \left[ (\delta\mu_c - \delta\mu_a)^2 
                + (\delta\mu_c - \delta\mu_b)^2 -2 (\delta\mu_a - \delta\mu_b)^2
               \right] 
                                                          \nonumber   \\
      & & + {\textstyle{3\over 2}} \tilde{C}_2
               (\delta\mu_c - \delta\bar{\mu})
                            (\delta m_u^2 + \delta m_d^2 + \delta m_s^2)
          + 6 (\tilde{C}_3 - \tilde{C}_4)
             ( \delta\mu_c - \delta\bar{\mu}) \delta\bar{\mu}^2 
                                                          \nonumber   \\
      & & + {\textstyle{1\over 6}} \tilde{Q}_3
            \left[ (\delta\mu_c - \delta\mu_a)^2 + 
                   (\delta\mu_c - \delta\mu_b)^2 
                   - 2 (\delta\mu_a - \delta\mu_b)^2 
            \right]\delta\bar{\mu}
                                                          \nonumber   \\
      & & + {\textstyle{1\over 8}} \tilde{Q}_4 
            (\delta\mu_c - \delta\bar{\mu})
               (\delta\mu_a^2 + \delta\mu_b^2 + \delta\mu_c^2 
               - 3\delta\bar{\mu}^2) \,,
                                                          \nonumber   \\
   \tilde{P}_{E^-} 
      &=& {\textstyle{\sqrt{3}\over 2}} \tilde{A}_2 (\delta\mu_b - \delta\mu_a) 
                                                          \nonumber   \\
      & & + {\textstyle{\sqrt{3}\over 2}}
                                    \tilde{B}_2 (\delta\mu_b^2 - \delta\mu_a^2)
          + {\textstyle{\sqrt{3}\over 4}} (\tilde{B}_3 - \tilde{B}_4) 
            \left[(\delta\mu_c - \delta\mu_b)^2 - (\delta\mu_c - \delta\mu_a)^2
            \right]
                                                          \nonumber   \\
      & & + {\textstyle{\sqrt{3}\over 2}}
                            \tilde{C}_2 (\delta\mu_b - \delta\mu_a)
                            (\delta m_u^2 + \delta m_d^2 + \delta m_s^2) 
          + 2 \sqrt{3} (\tilde{C}_3 - \tilde{C}_4)(\delta\mu_b - \delta\mu_a)
                                      \delta\bar{\mu}^2 
                                                          \nonumber   \\
      & & + {\textstyle{1\over 8\sqrt{3}}} \tilde{Q}_4
               (\delta\mu_b - \delta\mu_a)
                (\delta\mu_a^2 + \delta\mu_b^2 + \delta\mu_c^2 
                - 3\delta\bar{\mu}^2 ) 
                                                          \nonumber   \\
       & & - {\textstyle{\sqrt{3}\over 2}} \tilde{Q}_3
                (\delta\mu_b - \delta\mu_a)
                  (\delta\mu_c - \delta\bar{\mu})\delta\bar{\mu} \,,
                                                          \nonumber   \\
    \tilde{P}_{A_2}
       &=& \tilde{C}_9 (\delta\mu_c - \delta\mu_a)(\delta\mu_c - \delta\mu_b)
                               (\delta\mu_a - \delta\mu_b)  \,,
\label{twidPtoO3_NNLO}
\end{eqnarray} 
where
\begin{eqnarray} 
   \tilde{Q}_1 &\equiv & 2\tilde{C}_3 + \tilde{C}_5 + \tilde{C}_7
                                                          \nonumber   \\
   \tilde{Q}_2 &\equiv &  \tilde{C}_5 - \tilde{C}_6 + \tilde{C}_7 + \tilde{C}_8
                                                          \nonumber   \\
   \tilde{Q}_3 &\equiv &  4 (\tilde{C}_3 - \tilde{C}_4) + 3 (\tilde{C}_5 
                                                       - \tilde{C}_7)
                                                          \nonumber   \\
   \tilde{Q}_4 &\equiv &  2 (\tilde{C}_3 - \tilde{C}_4) + 3 (\tilde{C}_5 
                           -\tilde{C}_7) - 9 (\tilde{C}_6 + \tilde{C}_8) \,,
\end{eqnarray} 
and $\delta\bar{\mu} 
  \equiv {\textstyle{1\over 3}} (\delta\mu_a + \delta\mu_b + \delta\mu_c)$.
\begin{eqnarray}
   \tilde{M}_\Sigma^2(aab)
      &=& 1 + \tilde{A}_1(2\delta\mu_a + \delta\mu_b) 
                    + \tilde{A}_2(\delta\mu_b - \delta\mu_a)
                                                         \nonumber   \\
      & &             + \tilde{B}_0\delta m_l^2
                      + \tilde{B}_1(2\delta\mu_a^2 + \delta\mu_b^2)
                      + \tilde{B}_2(\delta\mu_b^2 - \delta\mu_a^2) 
                      + \tilde{B}_3(\delta\mu_b - \delta\mu_a)^2
                                                         \nonumber   \\
      & &             - 2\tilde{C}_0\delta m_l^3
                      + 6[ \tilde{C}_1(2\delta\mu_a + \delta\mu_b) 
                          + \tilde{C}_2(\delta\mu_b - \delta\mu_a)
                         ]\delta m_l^2
                                                         \nonumber   \\
      & &             + \tilde{C}_3(\delta\mu_a + \delta\mu_b)^3
                      + \tilde{C}_4(\delta\mu_a + \delta\mu_b)^2
                                          (\delta\mu_a - \delta\mu_b)
                                                         \nonumber   \\
      & &             + \tilde{C}_5(\delta\mu_a + \delta\mu_b)
                                          (\delta\mu_a - \delta\mu_b)^2
                      + \tilde{C}_6(\delta\mu_a - \delta\mu_b)^3 \,,
\label{MN_phys_cubic_NNLO}
\end{eqnarray}
and
\begin{eqnarray}
   \tilde{M}_\Lambda^2(aa^\prime b)
      &=& 1 + \tilde{A}_1(2\delta\mu_a + \delta\mu_b) 
            - \tilde{A}_2(\delta\mu_b - \delta\mu_a)
                                                         \nonumber   \\
      & &             + \tilde{B}_0\delta m_l^2
                      + \tilde{B}_1(2\delta\mu_a^2 + \delta\mu_b^2)
                      - \tilde{B}_2(\delta\mu_b^2 - \delta\mu_a^2) 
                      + \tilde{B}_4(\delta\mu_b - \delta\mu_a)^2
                                                         \nonumber   \\
      & &             - 2\tilde{C}_0\delta m_l^3
                      + 6[ \tilde{C}_1(2\delta\mu_a + \delta\mu_b) 
                          -\tilde{C}_2(\delta\mu_b - \delta\mu_a)
                         ]\delta m_l^2
                                                         \nonumber   \\
      & &             + \tilde{C}_3(\delta\mu_a + \delta\mu_b)^3
                      + (\tilde{C}_4-2\tilde{C}_3)
                         (\delta\mu_a + \delta\mu_b)^2
                                  (\delta\mu_b - \delta\mu_a)
                                                         \nonumber   \\
      & &             + \tilde{C}_7(\delta\mu_a + \delta\mu_b)
                                 (\delta\mu_b - \delta\mu_a)^2
                      + \tilde{C}_8(\delta\mu_b - \delta\mu_a)^3 \,.
\label{MLam_phys_cubic_NNLO}
\end{eqnarray}

Eq.~(\ref{Dsym}) is generalised to
\begin{eqnarray}
   D^{\rm sym}_{\Sigma\Lambda}
      &=& \tilde{A}_2 + \tilde{B}_2(\delta\mu_a+\delta\mu_b)
                                                 \label{Dsym_NNLO} \\
      &&+ 6\tilde{C}_2\delta m_l^2
        + (\tilde{C}_3-\tilde{C}_4)(\delta\mu_a+\delta\mu_b)^2
        - {\textstyle{1\over 2}}
            (\tilde{C}_6+\tilde{C}_8)(\delta\mu_b-\delta\mu_a)^2 \,.
                                                         \nonumber
\end{eqnarray}


\section{Tables}
\label{pq_mass_tables}


Table~\ref{Babc_masses} gives the PQ baryon masses when all three
valence quarks are different, while Table~\ref{Baab_masses} gives
the masses when two valence quarks are mass degenerate. The three
sea quark kappa values are $\kappa_l$ (twice) and $\kappa_s$, while the
valence quark values are $\kappa_a$, $\kappa_b$, $\kappa_c$.

\begin{table}[!htb]

\begin{footnotesize}
   \begin{center}
      \begin{tabular}{cccccc|ll}
         $\kappa_l$ & $\kappa_s$ & $\kappa_a$ & $\kappa_b$ & $\kappa_c$ & 
         $V$ & $M_H(abc)$ &$M_L(abc)$ \\ 
         \hline
    0.120900 & 0.120900 & 0.120900 & 0.120512 & 0.120000 & $32^3\times 64$ &  0.5698(31) &  0.5580(26)  \\
    0.120900 & 0.120900 & 0.121095 & 0.120512 & 0.120000 & $32^3\times 64$ &  0.5577(38) &  0.5433(31)  \\
    0.120900 & 0.120900 & 0.121095 & 0.120900 & 0.120000 & $32^3\times 64$ &  0.5300(45) &  0.5123(39)  \\
    0.120900 & 0.120900 & 0.121095 & 0.120900 & 0.120512 & $32^3\times 64$ &  0.4892(48) &  0.4777(43)  \\
    0.120900 & 0.120900 & 0.120900 & 0.120000 & 0.118000 & $32^3\times 64$ &  0.7278(26) &  0.7087(24)  \\
      \end{tabular}
   \end{center}
\caption{Baryon masses used with valence quark kappa values
         $\kappa_a \not= \kappa_b \not= \kappa_c$.}
\label{Babc_masses}
\end{footnotesize}

\end{table}

\begin{table}[!htb]

\begin{footnotesize}
   \begin{center}
      \begin{tabular}{cccccc|ll}
         $\kappa_l$ & $\kappa_s$ & $\kappa_a$ & $\kappa_b$ & $\kappa_c$ & 
         $V$ & $M_\Sigma(aab)$ &$M_\Lambda(aa^\prime b)$ \\ 
         \hline
 0.120900 & 0.120900 & 0.120000 & 0.120000 & 0.118000 & $32^3\times 64$ &  0.7789(23) &  0.7684(23) \\
 0.120900 & 0.120900 & 0.120000 & 0.120000 & 0.120000 & $32^3\times 64$ &  \multicolumn{2}{c}{0.6588(23)} \\
 0.120900 & 0.120900 & 0.120000 & 0.120000 & 0.120512 & $32^3\times 64$ &  0.6232(24) &  0.6290(26) \\
 0.120900 & 0.120900 & 0.120000 & 0.120000 & 0.120900 & $32^3\times 64$ &  0.5945(26) &  0.6058(29) \\
 0.120900 & 0.120900 & 0.120512 & 0.120512 & 0.120000 & $32^3\times 64$ &  0.5945(25) &  0.5892(28) \\
 0.120900 & 0.120900 & 0.120512 & 0.120512 & 0.120512 & $32^3\times 64$ &  \multicolumn{2}{c}{0.5564(27)} \\
 0.120900 & 0.120900 & 0.120512 & 0.120512 & 0.120900 & $32^3\times 64$ &  0.5247(30) &  0.5328(34) \\
 0.120900 & 0.120900 & 0.120900 & 0.120900 & 0.116000 & $32^3\times 64$ &  0.7811(34) &  0.7438(33) \\
 0.120900 & 0.120900 & 0.120900 & 0.120900 & 0.118000 & $32^3\times 64$ &  0.6739(33) &  0.6442(32) \\
 0.120900 & 0.120900 & 0.120900 & 0.120900 & 0.120000 & $32^3\times 64$ &  0.5435(32) &  0.5277(37) \\
 0.120900 & 0.120900 & 0.120900 & 0.120900 & 0.120512 & $32^3\times 64$ &  0.5031(35) &  0.4938(41) \\
 0.120900 & 0.120900 & 0.120000 & 0.120000 & 0.121095 & $32^3\times 64$ &  0.5806(31) &  0.5960(37) \\
 0.120900 & 0.120900 & 0.120512 & 0.120512 & 0.121095 & $32^3\times 64$ &  0.5086(36) &  0.5234(46) \\
 0.120900 & 0.120900 & 0.120900 & 0.120900 & 0.121095 & $32^3\times 64$ &  0.4502(55) &  0.4606(80) \\
 0.120900 & 0.120900 & 0.121095 & 0.121095 & 0.120000 & $32^3\times 64$ &  0.5226(65) &  0.5029(123) \\
 0.120900 & 0.120900 & 0.121095 & 0.121095 & 0.120512 & $32^3\times 64$ &  0.4838(77) &  0.4744(177) \\
 0.120900 & 0.120900 & 0.121095 & 0.121095 & 0.120900 & $32^3\times 64$ &  0.4519(42) &  0.4613(300) \\
 0.121040 & 0.120620 & 0.120620 & 0.120620 & 0.120620 & $32^3\times 64$ &  \multicolumn{2}{c}{0.5265(16)} \\
 0.121040 & 0.120620 & 0.120620 & 0.120620 & 0.121040 & $32^3\times 64$ &  0.4907(21) &  0.5014(30) \\
 0.121040 & 0.120620 & 0.121040 & 0.121040 & 0.120620 & $32^3\times 64$ &  0.4697(33) &  0.4547(43) \\
 0.121040 & 0.120620 & 0.121040 & 0.121040 & 0.121040 & $32^3\times 64$ &  \multicolumn{2}{c}{0.4267(50)} \\
 0.121095 & 0.120512 & 0.120512 & 0.120512 & 0.120512 & $32^3\times 64$ &  \multicolumn{2}{c}{0.5446(16)} \\
 0.121095 & 0.120512 & 0.120512 & 0.120512 & 0.121095 & $32^3\times 64$ &  0.4971(21) &  0.5054(31) \\
 0.121095 & 0.120512 & 0.121095 & 0.121095 & 0.120512 & $32^3\times 64$ &  0.4690(37) &  0.4510(58) \\
 0.121095 & 0.120512 & 0.121095 & 0.121095 & 0.121095 & $32^3\times 64$ &  \multicolumn{2}{c}{0.4140(61)} \\
 0.121145 & 0.120413 & 0.120413 & 0.120413 & 0.120413 & $32^3\times 64$ &  \multicolumn{2}{c}{0.5682(13)} \\
 0.121145 & 0.120413 & 0.120413 & 0.120413 & 0.121145 & $32^3\times 64$ &  0.5092(19) &  0.5239(23) \\
 0.121145 & 0.120413 & 0.121145 & 0.121145 & 0.120413 & $32^3\times 64$ &  0.4761(39) &  0.4507(65) \\
 0.121145 & 0.120413 & 0.121145 & 0.121145 & 0.121145 & $32^3\times 64$ &  \multicolumn{2}{c}{0.4016(89)} \\
 0.120900 & 0.120900 & 0.120900 & 0.120900 & 0.120900 & $32^3\times 64$ &  \multicolumn{2}{c}{0.4673(27)} \\
 0.121166 & 0.120371 & 0.120371 & 0.120371 & 0.120371 & $48^3\times 96$ &  \multicolumn{2}{c}{0.5730(26)} \\
 0.121166 & 0.120371 & 0.120371 & 0.120371 & 0.121166 & $48^3\times 96$ &  0.5083(35) &  0.5247(46) \\
 0.121166 & 0.120371 & 0.121166 & 0.121166 & 0.120371 & $48^3\times 96$ &  0.4680(66) &  0.4322(66) \\
 0.121166 & 0.120371 & 0.121166 & 0.121166 & 0.121166 & $48^3\times 96$ &  \multicolumn{2}{c}{0.3817(123)} \\
      \end{tabular}
   \end{center}
\caption{Baryon masses used with valence quark kappa values
         $\kappa_a = \kappa_b$.}

\label{Baab_masses}
\end{footnotesize}

\end{table}


\clearpage



\end{document}